\documentclass[journal]{IEEEtran}
\usepackage{cite}
\usepackage{soul}
\usepackage{pgfplots}
\usepackage{datatool}

\usepgfplotslibrary{units}
\usepgfplotslibrary{groupplots}
\usepgfplotslibrary{fillbetween}
\pgfplotsset{compat=1.15}
\usepackage[normalem]{ulem}
\usepackage{soul}
\newcommand{\stkout}[1]{\ifmmode\text{\sout{\ensuremath{#1}}}\else\sout{#1}\fi}

\usepackage{amsmath} 

\DeclareUnicodeCharacter{2113}{\ensuremath{\ell}}
\ifCLASSINFOpdf
\else
\fi

\usepackage{amsmath,amsfonts,amssymb,mathtools,bm}
\usepackage{siunitx}

\usepackage{subcaption}

\usepackage{hyperref}
\hypersetup{colorlinks=true,linkcolor=blue,citecolor=blue,urlcolor=black}
\usepackage{url}
\usepackage{orcidlink}
\begin{document}
\title{Physics-Informed Neural Network-Driven Sparse Field Discretization Method for Near-Field Acoustic Holography}

%Physics-Informed Neural Network on Compressive Equivalent Source for Near-Field Acoustic Holography

%\title{Curved Space Transform: a new Approach for\\ Acoustic Imaging and Sound Field Analysis}


\author{Xinmeng~Luan~\orcidlink{0009-0004-9283-5511},
        Mirco~Pezzoli~\orcidlink{0000-0003-1296-0992},~\IEEEmembership{Member,~IEEE}, 
        Fabio Antonacci,~\orcidlink{0000-0003-4545-0315}~\IEEEmembership{Member,~IEEE} 
        Augusto Sarti~\orcidlink{0000-0002-5803-1702}~\IEEEmembership{Member,~IEEE}
        % <-this % stops a space
%\thanks{M. Shell was with the Department
%of Electrical and Computer Engineering, Georgia Institute of Technology, Atlanta,
}


\markboth{Journal of \LaTeX\ Class Files,~Vol.~31, June~2023}%
{Shell \MakeLowercase{\textit{et al.}}: Bare Demo of IEEEtran.cls for IEEE Journals}



\maketitle

\begin{abstract}
% Nearfield acoustic holography...
We propose the Physics-Informed Neural Network-driven Sparse Field Discretization method (PINN-SFD), a novel self-supervised, physics-informed deep learning approach for addressing the Near-Field Acoustic Holography (NAH) problem. 
Unlike existing deep learning methods for NAH, which are predominantly supervised by large datasets, our approach does not require a training phase and it is physics-informed. 
% Unlike existing deep learning methods for NAH, which are predominantly supervised by large datasets, our approach is entirely driven by physics. 
The wave propagation field is discretized into sparse regions, a process referred to as field discretization, which includes a series of set of source planes, to address the inverse problem.
Our method employs the discretized Kirchhoff-Helmholtz integral as the wave propagation model. 
By incorporating virtual planes, additional constraints are enforced near the actual sound source, improving the reconstruction process. 
% This method allows for the estimation of the velocity field of equivalent sources and the reconstruction of pressure and velocity fields at virtual planes within the discretized domain. These fields are subsequently propagated to the hologram plane to yield a predicted pressure field. 
Optimization is carried out using Physics-Informed Neural Networks (PINNs), where physics-based constraints are integrated into the loss functions to account for both direct (from equivalent source plane to hologram plane) and additional (from virtual planes to hologram plane) wave propagation paths. Additionally, sparsity is enforced on the velocity of the equivalent sources.
% Furthermore, sparsity is promoted by introducing a regularization term for the equivalent source velocities in the loss functions during optimization.
% In contrast to conventional Compressive-Equivalent Source Method (C-ESM) that utilizes wave-superposition for wave propagation, PINN-SFD leverages a discretized KH integral and introduces field discretization to apply additional constraints near the actual sound source. 
Our comprehensive validation across various rectangular and violin top plates, covering a wide range of vibrational modes, demonstrates that PINN-SFD consistently outperforms the conventional Compressive-Equivalent Source Method (C-ESM), particularly in terms of reconstruction accuracy for complex vibrational patterns. Significantly, this method demonstrates reduced sensitivity to regularization parameters compared to C-ESM. 
% Moreover, we clarify that enhanced hologram pressure reconstruction does not necessarily equate to improved source velocity reconstruction, a distinction often overlooked in the NAH literature.
\end{abstract}

\begin{IEEEkeywords}
near-field acoustic holography, physics-informed neural network, sparse
field discretization method, compressive equivalent source method, kirchhoff-helmholtz integral
\end{IEEEkeywords}

% \fi
\IEEEpeerreviewmaketitle

\section{Introduction}\label{sec:introdution}

\IEEEPARstart{N}{ear-field} acoustic holography (NAH) is a widely used technique for identifying and visualizing sound sources in acoustics \cite{williams2000fourier}. 
% NAH can be broadly categorized as a sound field reconstruction problem, operating under the near-field assumption.
By analyzing the sound pressure data from a nearby microphone array (at the so-called hologram plane), NAH reconstructs surface velocity fields, allowing for the identification of vibrating regions and providing insights into how complex sources radiate into the medium. 
In addition to its diagnostic capabilities \cite{xu2024experimental}, NAH offers practical advantages, particularly for delicate or lightweight objects such as musical instruments \cite{le2008couchet}, as it avoids potential damage from accelerometers and the added mass of measuring equipment. 
% NAH is particularly advantageous for delicate or lightweight objects, such as musical instruments, as it avoids potential damage from accelerometers and the added mass of measuring equipment. 
% NAH is widely used in various applications, including noise control \cite{xu2024experimental, hou2013application} and musical instrument analysis \cite{le2012modal, goldsberry2012using}, among others.

Predicting surface velocity from hologram sound pressure, which involves inverting the Kirchhoff-Helmholtz (KH) integral, is a highly ill-conditioned process due to capturing the evanescent waves emitted by the sound source in near-field, thus it typically requires regularization \cite{williams2001regularization}.
Additionally, the problem is usually under-determined, meaning there are more surface points than measurements, leading to a non-unique solution subspace. 
The basic NAH problem addresses the scenario of stationary sound fields at a single frequency \cite{koopmann1989method, sarkissian2005method, valdivia2006study, bi2005nearfield,  chardon2012near, wang2018sparse, fernandez2017sparse, fernandez2018compressive}. Subsequently, transient NAH techniques have been developed to reconstruct time-dependent pressure and visualize the sound field in both time and space domains, moving toward real-time industrial applications \cite{hald2001time, geng2024reconstruction}.

Numerous methods exist for NAH, with some of the most popular including Fourier-based NAH \cite{williams1980sound, williams2000fourier}, plane or spherical wave expansions (such as the Helmholtz Least Squares (HELS) method \cite{wu1998reconstructing}, statistically optimized NAH \cite{jacobsen2007statistically}), and inverse numerical approaches (such as the inverse boundary element method \cite{bai1992application, kim1996reconstruction}).
% \textcolor{red}{TODO: add IBEM.}
Another commonly used technique is the Equivalent Source Method (ESM), also referred to as the wave superposition method \cite{koopmann1989method, sarkissian2005method, valdivia2006study, bi2005nearfield}. 
It is grounded in the Huygens–Fresnel principle, which posits that every point on a wavefront acts as a secondary source of spherical wavelets.
ESM models the sound emitted from a source surface using a layer of virtual monopole point sources positioned slightly inside the physical source, radiating into the free field \cite{koopmann1989method}. 
The method is based on the fundamental idea that an arbitrary wave-field can be expressed as the superposition of waves radiated by a collection of point sources. 
The weights of the equivalent sources are estimated by minimizing the error between the measured pressure field and the pressure field obtained by propagating the equivalent sources. 
Once the weights are estimated, the source velocity field is computed by propagating the equivalent sources to the actual source plane.

The classical optimization approach to solve inverse problems in NAH is through regularization in a least-squares sense, such as Tikhonov regularization, which uses the $\ell_2$-norm to promote smooth, minimum-energy estimates for the solution subspace \cite{williams2001regularization, kim2004optimal}.
Another approach, compressive sensing (CS), has been introduced into NAH with the goal of acquiring a sparse representation of the sound field \cite{chardon2012near}, and has been applied in various sound field reconstruction \cite{pezzoli2022sparsity, koyama2019sparse,verburg2018reconstruction, damiano2024zero} and NAH methods: Fourier-based NAH \cite{chardon2012near}, HELS \cite{wang2018sparse}, ESM \cite{fernandez2017sparse, fernandez2018compressive} and  Dictionary ESM \cite{canclini2017dictionary, de2022group}.
%for transient NAH :\cite{ geng2024reconstruction}. 
CS intuitively exploits the underlying sparse structure of the problem to achieve accurate signal reconstruction \cite{candes2008introduction}.
Sparse solutions are ideally achieved by solving the $\ell_0$-norm problem. However, this approach involves a combinatorial, non-convex search that quickly becomes computationally intractable. Conversely, compressed sensing (CS) suggests that the $\ell_0$-norm minimization can be relaxed to a convex $\ell_1$-norm minimization when the problem exhibits sparsity and the sensing matrix columns are sufficiently incoherent \cite{fernandez2017sparse}.
% The method that combines CS and ESM, is the C-ESM \cite{fernandez2017sparse}, where the source model is modeled by ESM and the inverse problem is solved by CS.
Later on, methods with the combination of $\ell_1$-norm and $\ell_2$-norm minimization were proposed \cite{rahimi2019scale, pham2017noise, wang2020accelerated, huang2020ratio}. 

% This objective can be achieved by combining compressive sensing (CS) technique and ESM, namely compresive-equivalent source method (C-ESM) \cite{fernandez2017sparse}.

% These approaches are primarily data-driven and supervised, relying on large datasets to train NNs effectively. 

With the rapid advancement of deep learning, it has gained significant attention in addressing acoustics problems, achieving notable success \cite{olivieri2020inter, olivieri2021eusipco, olivieri2021pinn, luancomplex, wang20213d, wang2022research, wang2023cylindrical, chaitanya2023machine, lobato2024using, miotello2023deep, caviedes2021gaussian, olivieri2023real, fernandez2023generative, karakonstantis2023generative, koyama2024physics, 
karakonstantis2024room, olivieri2024physics, bi2024point, miotello2024physics, morena2024reconstruction, ma2023physics}. 
Among these studies, several \cite{olivieri2020inter, olivieri2021eusipco, olivieri2021pinn, luancomplex, wang20213d, wang2022research, wang2023cylindrical, chaitanya2023machine, lobato2024using} specifically address NAH problem and all fall under the category of Physics-Guided Neural Networks (PGNNs).
According to the classification of Neural Network (NNs) for enforcing underlying physics, PGNNs are described as supervised, data-driven approaches \cite{Faroughi2024}. 
These frameworks construct surrogate mappings between well-formatted inputs and outputs, which are generated through controlled experiments and computations.  PGNNs require a large and sufficient dataset to be trained and used reliably.
In \cite{olivieri2021eusipco, olivieri2020inter, olivieri2021pinn, luancomplex, wang20213d, wang2022research, wang2023cylindrical}, NNs are typically employed to map hologram sound pressure to source velocity. 
For example, a complex-valued UNet-based convolutional neural network (CNN) (CV-KHCNN) was proposed in \cite{luancomplex}, building on earlier frameworks \cite{olivieri2021pinn, olivieri2021eusipco, olivieri2020inter}. Another example is a 3D CNN-based framework developed for NAH \cite{wang20213d, wang2022research, wang2023cylindrical}.
% The KH integral is also incorporated into the loss function, leading to the framework being named the Complex-Valued Kirchhoff-Helmholtz Convolutional Neural Network (CV-KHCNN).
On the other hand, rather than directly employing NNs to model the sound field data, some studies concentrate on learning the indirect intermediate variables utilized in traditional NAH methods. 
For instance, the method proposed in \cite{chaitanya2023machine} utilizes NNs to estimate the coefficients of the equivalent sources for ESM. In \cite{lobato2024using}, an invertible NN \cite{ardizzone2018analyzing} is utilized together with HELS method to improve the inversion results.

Another deep learning framework, Physics-Informed Neural Networks (PINNs) \cite{raissi2019physics} have found extensive applications in acoustics, including sound field reconstruction \cite{karakonstantis2024room, olivieri2024physics, bi2024point}, upsampling \cite{miotello2024physics}, directivity modeling \cite{morena2024reconstruction}, and Head-Related Transfer Function (HRTF) estimation \cite{ma2023physics}.
According to the classification of NNs for incorporating physical laws \cite{Faroughi2024}, PINNs respect these laws by employing weakly imposed loss functions based on residuals of the governing physics equations, boundary and initial conditions. The networks typically take spatial and temporal coordinates as inputs and produce the desired physical quantities as outputs. By utilizing automatic differentiation, PINNs compute the necessary derivatives to evaluate these residuals and minimize the loss function, thereby approximating solutions to the underlying physical systems. As a result, PINNs function as a self-supervised framework, eliminating the reliance on large labeled datasets and instead leveraging the governing equations as a physics-informed regularization mechanism.
% This makes PINNs a self-supervised approach, eliminating the need for large training datasets while leveraging knowledge of governing equations as a physics-informed regularization mechanism.
% computational mathematics and physics, achieving significant success \cite{ray2024deep}. They are applicable to both forward problems (solving equations) and inverse problems (recovering boundary/initial conditions or equation coefficients from observations) \cite{ray2024deep}. Of particular importance are inverse problems, which hold immense potential in engineering applications due to their robustness against noisy real-world data \cite{ray2024deep}.
The aforementioned CV-KHCNN \cite{luancomplex} and its earlier version KHCNN \cite{olivieri2021pinn} exemplify approaches that integrate data-driven modeling with physical constraints. These frameworks incorporate the KH integral into their loss functions to enforce wave propagation physics, aligning with the principles of PINNs. Consequently, CV-KHCNN and KHCNN can be regarded as hybrids of PGNNs and PINNs. Unlike most PINN-based methods that solve PDEs, however, CV-KHCNN and KHCNN focus on addressing the KH integral.

% PINNs are being explored for sound field reconstruction \cite{koyama2024physics, olivieri2021pinn, luancomplex}, also for applications in NAH \cite{luancomplex, olivieri2021pinn}.

% CV-KHCNN \cite{luancomplex, olivieri2021PINN} was inspired by the paradigm of PINNs \cite{raissi2019physics}. Building on the model in \cite{olivieri2021eusipco, olivieri2020inter}, the authors proposed to incorporate the KH integral into the loss functions. However, even with the physics constraint in the loss function, the CNN is still trained on a large dataset. In other word, the solution of the NNs is still mostly driven by large data than the physical law.

% CV-KHCNN provides two outputs: source velocity and sound pressure, which are also used to calculate the KH integral loss. However, due to the normalization of inputs (hologram pressure) and outputs in CV-KHCNN to match the operating range of the NNs, the KH integral may not be computed at the correct scale. As the KH integral is a nonlinear system, this normalization mismatch can prevent the loss function from operating in the expected scale.

As mentioned, existing deep learning approaches for NAH rely on supervised learning; however, NAH has broad applications, including noise control \cite{xu2024experimental, hou2013application} and musical instrument analysis \cite{le2012modal, goldsberry2012using, pezzoli2022comparative, de2022group}, among others. The limitation of these supervised methods \cite{olivieri2020inter, olivieri2021eusipco, olivieri2021pinn, luancomplex, wang20213d, wang2022research, wang2023cylindrical, chaitanya2023machine, lobato2024using} restricts their usefulness to sound sources similar to those in the training data. While training with large models on extensive datasets might alleviate this limitation, such comprehensive datasets are currently unavailable. 

With the objective to explore a more adaptable approach, we propose a physics-informed deep learning method for the NAH problem: the Physics-Informed Neural Network-Driven Sparse Field Discretization (PINN-SFD) method, governed by the KH Integral.
% Specifically, the PINN-SFD operates in a one-shot, self-supervised manner: \textit{one-shot} indicates that the optimization problem is based on a single sample without necessitating a large training dataset, while \textit{self-supervised} signifies that the optimization is driven by the physics-informed regularization agent, eliminating the need for ground truth data.
% Firstly, PINN-SFD is a self-supervised approach, where only physics constraints driven the optimization problem. 
We adopt a similar approach to ESM by estimating the velocity field of the equivalent sources.
% Unlike conventional ESM approaches \cite{koopmann1989method}, which use wave-superposition for wave propagation, 
In addition, we introduce the concept of Field Discretization (FD), which applies extra constraints near the actual sound source. Once the equivalent sources are determined, they propagate to Virtual Planes (VPs) positioned between the equivalent source plane and the hologram plane. This facilitates the reconstruction of both velocity and pressure fields at these VPs. These reconstructed fields then serve as secondary sources, which are further propagated to the hologram plane, resulting in an additional predicted pressure at the hologram plane.
Wave propagation is modeled using the discretized KH integral.
The optimization problem is then addressed using PINNs, where all physics constraints are embedded into the loss functions. These constraints capture both the direct propagation path from the equivalent source plane to the hologram plane, as well as the additional paths through VPs. Additionally, a regularization term is applied to impose sparsity in the equivalent sources velocity.
In the proposed PINN-SFD method, \textit{Field Discretization} refers to the discretization of the wave propagation field through both the equivalent source plane and the virtual planes, while \textit{Sparse} reflects the sparsity imposed both in the discretized field and the optimization process.
Specifically, the PINN-SFD operates in a ``single-instant'', ``self-supervised manner'': \textit{single-instant} indicates that the optimization problem is based on a single sample without necessitating a large training dataset, while \textit{self-supervised} signifies that the optimization is driven by the physics-informed regularization agent, eliminating the need for ground truth data.
% PINN-CESM is a one-shot method, similar to inverse PINNs. 
% However, instead of solving the governing PDE, specifically the Helmholtz equation for wave propagation directly, we still focus on the boundary integral solution to the Helmholtz equation, i.e., the KH integral, as previous \cite{luancomplex, olivieri2021pinn}.
% The wave propagation model in the loss function is similar to C-ESM, where we also introduce the equivalent sources to aid the source reconstruction. 
% While direct discrete KH integral is utilized for propagation in this paper, instead of the wave-superposition model in conventional C-ESM. Leveraging the KH integral’s ability to propagate fields independently, we propose the idea of additional Virtual Planes (VPs), which aims to impose additional constraints near the real source. This is the so-called PINN-VP-CESM. Therefore, the loss function includes three components: the sound pressure loss at the hologram plane from both the direct path (equivalent sources to hologram plane) and the indirect path (virtual planes to hologram plane), as well as a regularization term for sparsity.

The proposed PINN-SFD is validated across multiple rectangular plates and violin top plates, encompassing a wide range of vibrational modes. 
The results show that PINN-SFD offers more consistent performance than conventional C-ESM across different vibrational patterns, making it a reliable choice for source field reconstruction tasks. 
% Particularly, unlike the conventional C-ESM, PINN-SFD is less sensitive to regularization parameters. Moreover, the accuracy of the reconstruction is highly related to the complexity of the vibrational pattern, demonstrating high reconstruction ability for complex patterns while struggling with overly simple or complex ones. An interesting aspect is that VPs improve the reconstruction of fine details, especially in highly complex vibrational patterns. We also address the issue that better hologram pressure reconstruction does not necessarily translate to better source velocity reconstruction, which is often overlooked in NAH scenario.
Unlike conventional C-ESM, PINN-SFD is less sensitive to regularization parameters. It excels at reconstructing complex vibrational patterns but faces challenges with overly simple or highly intricate ones. 
Additionally, we highlight that VPs are more efficient for high-frequency complex patterns, enhancing fine detail reconstruction. However, when tracking the accuracy of the actual source velocity along training epochs, we observe that the rebound effect for the actual source velocity may occur in  low-frequency modes when utilizing VPs.
The implementation is available on GitHub \footnote{\href{https://github.com/Xinmeng-Luan/PINN-SFD}{https://github.com/Xinmeng-Luan/PINN-SFD}}.

The remainder of this paper is organized as follows. Section \ref{sec:signal_model_background} describes the 
KH integal and C-ESM. Section \ref{sec:method} describes PINN-SFD. Section \ref{sec:validation} presents the results and discussion of PINN-SFD. Finally, Section~\ref{sec:conclusion} summarizes the study and outlines future directions.

\section{Signal Model and Background}\label{sec:signal_model_background}
%===================================================
\subsection{NAH problem formulation}

The goal of NAH is to reconstruct the normal velocity of a sound source from measured near-field pressure data \cite{williams2000fourier}, typically by solving the inverse Kirchhoff-Helmholtz (KH) integral. The KH integral offers a solution to the inhomogeneous Helmholtz equation for radiation problems in exterior domains, while satisfying the Sommerfeld radiation condition \cite{williams2000fourier, atkinson1949lxi}. 

\begin{figure}
    \centering
        \includegraphics[width=0.8\linewidth]{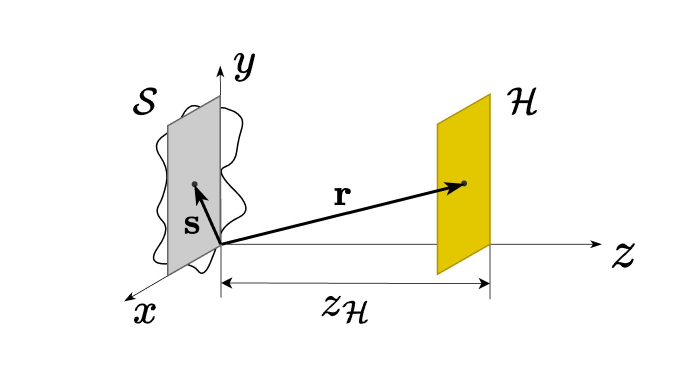}
        \caption{General setup for NAH. }
        \label{fig:nah-setup}
\end{figure}

For instance, consider a vibrating surface, $\mathcal{S}$, such as the top plate of a string musical instrument, with points at $\mathbf{s} \in \mathbb{R}^3$, and a near-field plane, referred to as the hologram plane $\mathcal{H}$, containing sound pressure measurement points $\mathbf{r} \in \mathbb{R}^3$, as illustrated in Fig.~\ref{fig:nah-setup}.
% The exterior pressure field radiated from a vibrating structure can be characterized by the well-known Kirchhoff-Helmholtz (KH) integral \cite{williams2000fourier}. 
% NAH is typically solved using the well-known Kirchhoff-Helmholtz (KH) integral, which characterizes the exterior pressure field radiated by a vibrating structure \cite{williams2000fourier}.
% The inhomogeneous Helmholtz equation is 
% \begin{equation}
%     \nabla^2 p(\mathbf{r}, \omega) + k^2 p(\mathbf{r}, \omega) = -Q(\mathbf{r}),
% \end{equation}
% where $\displaystyle{p(\mathbf{r}, \omega) = \int_\Omega g_\omega (\mathbf{r}, \mathbf{s}, \omega) Q(\mathbf{s}) d\Omega'}$.
% There are two Boundary Conditions (BCs) need to consider for the exterior domain radiation problem, 
% two boundary condition:
 The forward problem, which characterizes the acoustic field generated at $\mathcal{H}$  by the sound source located at $\mathcal{S}$, can be addressed using the KH integral, as \cite{williams2000fourier}
% \begin{equation} \label{eq:kirchhoffhelmholtz}
% \begin{aligned}
%     p(\mathbf{r}, \omega) &= \int_\mathcal{S} p(\mathbf{s}, \omega) \frac{\partial}{\partial \mathbf{n}} g_{\omega}(\mathbf{r},\mathbf{s}) d\mathbf{s} \\ 
%     &- j \omega \rho_0 \int_\mathcal{S} v_{n}(\mathbf{s},\omega) g_{\omega}(\mathbf{r},\mathbf{s}) d\mathbf{s},
% \end{aligned}
% \end{equation} 
\begin{align} \label{eq:kh}
    \alpha(\mathbf{r}) p(\mathbf{r}, \omega) = \int_\mathcal{S} \bigg( 
    & p(\mathbf{s}, \omega) \frac{\partial}{\partial \mathbf{n}} g_{\omega}(\mathbf{r},\mathbf{s}) \notag \\
    & - g_{\omega}(\mathbf{r},\mathbf{s}) \frac{\partial}{\partial \mathbf{n}} p(\mathbf{s},\omega) \bigg) d \mathcal{S},
\end{align}
where $\mathbf{n} \in \mathbb{R}^{3}$ is the outward normal direction unit vector, $p(\cdot,\omega)$ is the pressure, and $\omega$ is the angular frequency.
$g_{\omega}(\mathbf{r},\mathbf{s})$ is the free-field Green's function from $\mathbf{s}$ to $\mathbf{r}$, written as \cite{williams2000fourier}
\begin{equation} \label{eq:greenfunction}
    g_{\omega}(\mathbf{r},\mathbf{s}) = \frac{1}{4\pi}\frac{e^{-jk\left | \left |\mathbf{r}-\mathbf{s}\right | \right |}}{\left | \left |\mathbf{r}-\mathbf{s} \right | \right |},
\end{equation}
with $c$ the sound speed in the air, $j$ the imaginary unit, and $k=\omega/c$ the wave number. The parameter $\alpha(\mathbf{r})$ in \eqref{eq:kh} is determined by the position of the measured point $\mathbf{r}$:
\begin{equation}
   \alpha(\mathbf{r}) = 
   \begin{cases}
        1, & \text{if $\mathbf{r}$ is outside $\mathcal{S}$}, \\
        1/2, & \text{if $\mathbf{r}$ is on $\mathcal{S}$}, \\
        0,  & \text{if $\mathbf{r}$ is inside $\mathcal{S}$}.
    \end{cases}
\end{equation}
% Moreover, the Sommerfeld radiation condition is satisfied to derive (\ref{eq:kh})  \cite{williams2000fourier, atkinson1949lxi}, which gives a Boundary Condition (BC) at infinity, as 
% \begin{equation}
%     \lim_{\mathbf{r} \to \infty} \mathbf{r} \left [ \frac{\partial}{\partial \mathbf{n}} p(\mathbf{r}) - j k p(\mathbf{r})\right] = 0.
% \end{equation}
Furthermore, the relation between the pressure and the normal velocity is characterized by Euler's equation \cite{williams2000fourier}
\begin{equation}\label{eq:euler}
    \frac{\partial}{\partial \mathbf{n}} p(\mathbf{s},\omega) = j\omega \rho v(\mathbf{s},\omega),
\end{equation}
where $\rho \approx \SI{1.225}{\kilo\gram\per\cubic\meter}$  is the air mass density at $\SI{20}{\celsius}$ and $v(\cdot,\omega)$ is the normal velocity field. 
Note that the explicit form of the KH integral used to compute the generated sound pressure and velocity field is derived in Appendix \ref{sec:appendix1}. 

% In practical applications, only discretized data can be processed. Since this study focuses on the planar NAH scenario,  the derivation of the discrete KH integral for the planar NAH configuration is provided in Appendix \ref{sec:appendix2}. For simplicity, denote discrete KH integral as operators: \eqref{eq:ds-ps} as $\Theta_1$, 
% \eqref{eq:ds-ph} as $\Theta_2$, and
% \eqref{eq:ds-vh} as $\Theta_3$. 

% \begin{figure}
%     \centering
%     \begin{minipage}{0.6\linewidth}
%         \centering
%         \includegraphics[width=\linewidth]{figs/general/NAH_config.png}
%     \end{minipage}%
%     % \hfill
%     \hspace{1ex}
%     \begin{minipage}{0.25\linewidth}
%         \caption{General setup for NAH. }
%         \label{fig:nah-setup}
%     \end{minipage}
% \end{figure}

% Using the KH integral, we can calculate the pressure emitted by a vibrating source based on the pressure and normal velocity fields at the surface of the object. 
The inversion of this forward problem, i.e., NAH, can be mathematically expressed as
\begin{equation}
    \hat{v}(\mathbf{s},\omega) \Big |_{\mathbf{s} \in \mathcal{S}} \approx \Lambda^{-1} \left [ p(\mathbf{r},\omega)\right] \Big | _{\mathbf{r} \in \mathcal{H}},
\end{equation}
where $\Lambda$ is a estimator that approximates the pressure field on $\mathcal{H}$ given the normal velocity field on $\mathcal{S}$. 
However, this inverse problem is a highly ill-posed process as the free-field Green's function  represents the evanescent wave, which models the decay of spherical waves with distance, and its inversion necessitates regularization \cite{williams2001regularization}. The challenge becomes even greater in under-determined scenarios where the resolution of hologram measurements is lower than the desired source resolution, a common occurrence in real-world situations, leading to a non-unique solution subspace.

%%%%%%%%%%%%%%%%%%%%%%%%%%%%%%%%%%%%%%%%%%%%%%%%%%%%%%%%%%%%%%%%%%%%%%%%%%%%%%%%%%%%%%%%%%%%%%%%%%%%%%%%%%%%%%%%%%%%%%
\subsection{Equivalent Source Method in NAH}\label{sec: esm}

The ESM \cite{sarkissian2005method, koopmann1989method, valdivia2006study, bi2005nearfield} models the sound emitted from a source surface using a layer of virtual equivalent monopole point sources positioned slightly inside the physical source, radiating into the free field. 
Determination of the complex equivalent source amplitudes typically relies on sound pressure data measured by a nearby microphone array. 
% The ESM is also known as the method of wave superposition or source simulation method \cite{koopmann1989method}. 
% The method is based on the fundamental idea that an arbitrary wave-field can be expressed as the superposition of waves radiated by a collection of point sources. 
% The weights of the equivalent sources are estimated by minimizing the error between the measured pressure field  and the pressure field obtained by propagating the equivalent sources.
Instead of directly solving the problem through the KH integral, ESM is based on the fundamental idea that an arbitrary wave-field can be expressed as the superposition of waves radiated by a collection of point sources \cite{koopmann1989method}. 
The ESM is also known as the single layer potential method \cite{williams2000fourier}.
Let $\mathcal{E}$ denote equivalent source plane with points at $\mathbf{s'}$. The generated exterior measured sound pressure and normal velocity at $\mathbf{r}$ from the equivalent sources can be expressed by
% \begin{equation}\label{eq:wave_sup_}
% \begin{cases}
%       \displaystyle{p(\mathbf{r},\omega)= j \omega \rho \int_\mathcal{E} q(\mathbf{s'}) g_{\omega}(\mathbf{r},\mathbf{s'}) d\mathcal{E},} \\
% \displaystyle{v(\mathbf{r},\omega)= - \int_\mathcal{E} q(\mathbf{s'}) \frac{\partial}{\partial \mathbf{n}} g_{\omega} 
%       (\mathbf{r},\mathbf{s'})  d\mathcal{E},}
% \end{cases}
% \end{equation}

\begin{equation}\label{eq:wave_sup}
\begin{cases}
      \displaystyle{p(\mathbf{r},\omega) = j \omega \rho \sum_{n=1}^{N_1} q(\mathbf{s'}_n) \, g_{\omega}(\mathbf{r},\mathbf{s'}_n) \, \Delta \mathcal{E}_n,} \\
      \displaystyle{v(\mathbf{r},\omega) = - \sum_{n=1}^{N_1} q(\mathbf{s'}_n) \, \frac{\partial}{\partial \mathbf{n}} g_{\omega}(\mathbf{r},\mathbf{s'}_n) \, \Delta \mathcal{E}_n,}
\end{cases}
\end{equation}
where $q(\mathbf{s'})$ is the equivalent source volume velocity and $N_1$ is the number of virtual sources. This wave propagation based on the superposition \eqref{eq:wave_sup} has been proven to be equivalent to the KH integral  \eqref{eq:kh} in \cite{koopmann1989method}.

% In practice, we consider discrete systems. For the NAH scenario, the sound pressure captured by $M$ microphones in the hologram plane
In practice, the hologram plane is sampled with $M$ microphones, while the actual source, described over $N_2$ spatial grid points, can be equivalently represented through $N_1$  virtual sources positioned behind the real source.
% we sample the hologram plane using $M$ microphones, and the desired real source with $N_2$ spatial grids can be equivalently characterized by $N_1$ virtual sources retracted behind the real source. 
Then the pressure can be discretized into matrix form, defined as 
\begin{equation}
\mathbf{p}_\mathcal{H}=\mathbf{G}_\mathcal{H}{\mathbf{q}},
\label{eq:matrix-cesm-h}\end{equation}
where the subscript $\mathcal{H}$ represents the hologram plane. $\mathbf{p}_\mathcal{H} = [p(\mathbf{r_1},\omega), ..., p(\mathbf{r}_M,\omega)]^T \in \mathbb{C}^M$ is the measured pressure, ${\mathbf{q}}= [q_1, ..., q_{N_1}]^T \in \mathbb{C}^{N}$ is the coefficient vector containing the strength of the sources, which are related to their volume velocity $\mathbf{q'}$ as $\mathbf{q}=j\omega \rho \mathbf{q'}$, and $\mathbf{G}_\mathcal{H} \in \mathbb{C}^{M\times N_1}$ represents the Green's function pairs describing the wave propagation between the positions of the equivalent sources and microphones. 
% The solution of $\eqref{eq:matrix-cesm-h}$ leads to an estimate $\hat{\mathbf{q}}$. 

Therefore, the solution to the NAH problem through ESM becomes to find the ESM coefficient vector $\hat{\mathbf{q}}$ by inverting \eqref{eq:matrix-cesm-h}, where the solution is found by solving an optimization problem. Note that $\hat{\cdot}$ refers to predicted quantities in this paper.
% This problem is highly ill-conditioned since the free-field Green's function models the decay of spherical waves over distance, and its inversion necessitates regularization. The challenge becomes even greater in under-determined scenarios where $N_1 > M$, a common occurrence in real-world situations, leading to a non-unique solution subspace.
The regularization for solving the system of \eqref{eq:matrix-cesm-h} is formulated as 
\begin{equation}
    \hat{\mathbf{q}} = \arg_{\mathbf{q}} \min \left( \| \mathbf{p}_\mathcal{H} - \mathbf{G}_\mathcal{H} \mathbf{q} \|_a^a + \lambda \| \mathbf{q} \|_b^b \right),
    \label{eq:cesm-formulation}
\end{equation}
where $\lambda>0$  is the regularization parameter that balances the trade-off between the data fitting term, represented by the $l_a$-norm (commonly with $a=2$), which is the first term, and the regularization term, indicated by the $l_b$-norm (the second term), which imposes a penalty on the solution.
Compressive-equivalent source method (C-ESM) utilizes $l_1$-norm for the minimization problem in our case, referring to the penalty term where ($b=1$), which inherently promotes a sparse solution \cite{fernandez2017sparse, gade2016wideband, chardon2012near, abusag2016sparse, hald2016fast, ping2017refined}.
The most popular algorithm to solve the inverse problem is the interior-point convex optimization algorithm implemented in the publicly available MATLAB toolbox CVX \cite{grant2014cvx}, which is used in \cite{fernandez2017sparse, hald2016fast, abusag2016sparse, chardon2012near, ping2017refined}. For additional algorithms, please refer to \cite{hald2016fast, ping2017refined, suzuki2011l1, xu2014high, oudompheng2014theoretical, beck2009fast, rish2014sparse, hald2018comparison} for further insights.

Once $\hat{\mathbf{q}}$ is solved, subsequently the velocity at the actual source surface can be reconstructed by
\begin{equation}
    \hat{\mathbf{v}}_\mathcal{S} =- \frac{1}{j \omega \rho} \frac{\partial \mathbf{G}_\mathcal{S} }{\partial n}\hat{\mathbf{q}},
    % \quad \hat{\mathbf{p}}_\mathcal{S}=\mathbf{G}_\mathcal{S}\hat{\mathbf{q}},
\end{equation}
where the subscript $\mathcal{S}$ represents the real source surface plane.
$\hat{\mathbf{v}}_\mathcal{S} = [\hat{v}(\mathbf{s_1},\omega), ..., \hat{v}(\mathbf{s}_{N_2},\omega)]^T \in \mathbb{C}^{N_2}$ is the reconstructed actual source velocity, 
$\hat{\mathbf{p}}_\mathcal{S} = [\hat{p}(\mathbf{s_1},\omega), ..., \hat{p}(\mathbf{s}_{N_2},\omega)]^T \in \mathbb{C}^{N_2}$ is the reconstructed surface pressure, and $\mathbf{G}_\mathcal{S} \in \mathbb{C}^{M\times N_2}$ represents the Green's function pairs between the positions of the equivalent sources and real source. 
%% FABIO: ARRIVED HERE

\section{Proposed method: PINN-SFD}\label{sec:method}
% \subsection{PINN-FD-SESM Overview}
% We now propose a novel framework for solving the NAH problem: the Physics-Informed Neural Network-driven Sparse Field Discretization method (PINN-SFD), governed by the Kirchhoff-Helmholtz Integral.
\subsection{Optimization Problem Formulation}
% \textcolor{red}{This concept aligns with the approach used in PINNs when solving PDEs, where boundary and initial condition losses are incorporated into the loss function, a common strategy for optimizing PINNs.
% }

\begin{figure}
    \centering
\includegraphics[width=\linewidth]{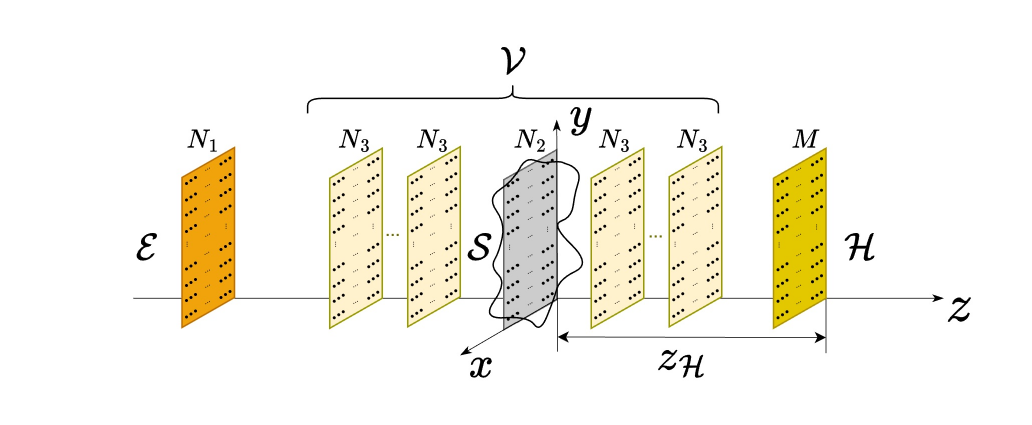}
    \caption{Planar NAH configuration for PINN-SFD, with $\mathcal{E}$ (equivalent source), $\mathcal{V}$ (virtual), $\mathcal{S}$ (actual source), and $\mathcal{H}$ (Hologram) planes.}
    \label{fig:CESM_config}
\end{figure}

Within the ESM framework, where a virtual equivalent source plane positioned behind the actual source is used for reconstruction, it is intuitive to consider that imposing additional constraints closer to the source plane may improve reconstruction accuracy.
In other words, when solving for the coefficient vector $\hat{\mathbf{q}}$ in \eqref{eq:cesm-formulation}, the optimization problem only considers the constraints on the relationship between the equivalent sources plane $\mathcal{E}$ and the measured pressure at the hologram plane $\mathcal{H}$ (see Fig.~\ref{fig:CESM_config}). However, as our primary goal is to reconstruct the velocity field at the actual source plane $\mathcal{S}$, it would be advantageous to apply constraints closer to this plane. To achieve this, we discretise the spatial domain during wave propagation (as $\mathcal{V}$ shown in Fig.~\ref{fig:CESM_config}), focusing on the region between the equivalent sources and the hologram plane, in proximity of the actual source plane $\mathcal{S}$, to incorporate additional constraints. These additional planes are referred to as Virtual Planes (VPs). Consequently, the wave propagation field is discretized into sparse regions, including the equivalent source plane $\mathcal{E}$ and VPs $\mathcal{V}$. With this approach, once the equivalent sources are predicted, they can propagate to the VPs, enabling the reconstruction of velocity and pressure fields at these VPs. These reconstructed fields then act as secondary sources, propagating further to the hologram plane and generating an additional predicted pressure at the hologram plane. The configuration of this proposed framework is shown in Fig.~\ref{fig:CESM_config}. 

To incorporate additional constraints and improve the reconstruction process, we directly employ the discretized KH integral as wave propagation model. 
%The derivation of the discretized KH integral in planar coordinates is presented as follows.
Assume there are two planes, positioned parallel to each other along the $z$-axis, as shown in Fig.~\ref{fig:nah-setup}. The coordinates of points on the first plane (source plane $\mathcal{S}$) are denoted by $(n_x, n_y, n_z)$, and the coordinates of points on the second plane (hologram plane $\mathcal{H}$) are denoted by $(m_x, m_y, m_z)$. 
There are $N= N_x \times N_y$ and $M= M_x \times M_y$ points on $\mathcal{S}$ and $\mathcal{H}$, respectively.
The distance between the two planes along the $z$-axis is $d_z = |n_z - m_z|$.
The discretization steps on $\mathcal{S}$ in the $x$- and $y$- directions are $\Delta L_{x}$ and $\Delta L_{y}$, respectively.
The distance between two points $\mathbf{s}$ and $\mathbf{r}$ on $\mathcal{S}$ and $\mathcal{H}$ is  $d = \|\mathbf{s}-\mathbf{r}\|_2$.
The Green's function (\ref{eq:greenfunction}) between $\mathbf{s}$ and $\mathbf{r}$, its first- and second-order derivatives  are
\begin{equation}
\begin{cases}
    g_\omega = \displaystyle \frac{e^{-jkd}}{4\pi  d},
    \\[1.5ex]
   \displaystyle \frac{\partial }{\partial \textbf{n}} g_\omega =  \displaystyle \frac{- e^{-jkd}(1+jkd)d_z}{4 \pi d^3},
   \\[1.5ex]
    \displaystyle  \frac{\partial ^2}{\partial \textbf{n}^2} g_\omega = \displaystyle \frac{e^{-jkd}(-k^2d+3+3jkd)d_z}{4\pi d^3},
     \end{cases}
     \label{eq:ds-green}
\end{equation}
The pressure field on $\mathcal{H}$ can be computed following (\ref{eq:kh2}). For the specific case where $\mathcal{H}$ coincide with $\mathcal{S}$, (\ref{eq:kh2}) is discretized as 
\begin{equation}\label{eq:ds-ps}
\begin{aligned}
& p(m_x,m_y, \omega) = \\
&  \qquad \begin{cases}
         \displaystyle \sum_{n_x=1}^{N_x} \sum_{n_y=1}^{N_y} -  \frac{1}{2 \pi} \frac{\omega^2}{c}\rho_0 v(n_x,n_y,\omega) \Delta L_{x} \Delta L_{y},  \\
 \qquad  \qquad  \qquad  \qquad  \qquad       \text{if $m_x = n_x$ and $m_y = n_y$,} \\
         \displaystyle \sum_{n_x=1}^{N_x} \sum_{n_y=1}^{N_y} -  \frac{1}{2 \pi} j\omega \rho _0 v (n_x,n_y,\omega) \frac{e^{-j k d}}{d}  \Delta L_{x} \Delta L_{y},  \\  \qquad  \qquad  \qquad  \qquad  \qquad  \qquad \qquad  \qquad  
 \text{otherwise,}
\end{cases}
\end{aligned}
\end{equation}
which is noted as $\Theta_1(v, n_x, n_y, d, \omega)$.

When $\mathcal{H} \neq \mathcal{S}$, 
\begin{equation}
\begin{aligned}
 p(m_x, & m_y, \omega) = \\ &
         \sum_{n_x=1}^{N_x} \sum_{n_y=1}^{N_y} \bigg[
         p(n_x,n_y, \omega) \displaystyle \frac{- e^{-jkd}(1+jkd)d_z}{4 \pi d^3}  \\
       & \quad  -  \displaystyle \frac{1}{4 \pi} j\omega \rho _0 v (n_x,n_y,\omega) \frac{e^{-j k d}}{d} \bigg] \Delta L_{x} \Delta L_{y},   \end{aligned}
\label{eq:ds-ph}
\end{equation}
which is noted as $\Theta_2(p, v, n_x, n_y, d, d_z, \omega)$.
The velocity field of $\mathcal{H}$ can be computed following  (\ref{eq:khv}), which is discretized as 
\begin{equation}
\begin{aligned}
   & v(m_x,m_y,\omega) = \\ &\sum_{n_x=1}^{N_x} \sum_{n_y=1}^{N_y}  
      \bigg[ \frac{1}{j\omega \rho_0} p(n_x,n_y, \omega) 
    \frac{e^{-jkd}(-k^2d+3+3jkd)d_z}{4\pi d^3} \\  & \qquad 
    - v(n_x,n_y,\omega) 
    \frac{- e^{-jkd}(1+jkd)d_z}{4 \pi d^3}
    \bigg] \Delta L_{x} \Delta L_{y},
    \end{aligned}
    \label{eq:ds-vh}
\end{equation}
and noted as $\Theta_3(p, v, n_x, n_y, d, d_z, \omega)$.
In the following, for clarity, we use more general representations $\Theta_1(\mathbf{v}, \mathbf{s}, \omega)$, $ \Theta_2(\mathbf{p}, \mathbf{v}, \mathbf{s}, \mathbf{r}, \omega), $ and $ \Theta_3(\mathbf{p}, \mathbf{v}, \mathbf{s}, \mathbf{r}, \omega)$. 

Define $\mathcal{V}_i$ as the $i$-th virtual plane, consisting of $N_3$ spatial grids at $\mathbf{s}''_i$, with $i \in [1, N_v]$. Here, the subscript represents the corresponding field, while the superscript indicates the originating field. Let $\hat{\mathbf{v}}_\mathcal{E} = [\hat{v}(\mathbf{s}_1', \omega), \dots, \hat{v}(\mathbf{s}_{N_1}', \omega)]^T \in \mathbb{C}^{N_1}$ denote the reconstructed equivalent source velocity, and note that $\hat{\mathbf{p}}_\mathcal{E}$ has the same dimension as $\hat{\mathbf{v}}_\mathcal{E}$. Furthermore, define $\hat{\mathbf{p}}_{\mathcal{V}_i}^{\mathcal{E}} = [\hat{p}(\mathbf{s}_1'', \omega), \dots, \hat{p}(\mathbf{s}_{N_3}'', \omega)]^T \in \mathbb{C}^{N_3}$, and note that $\hat{\mathbf{v}}_{\mathcal{V}_i}^{\mathcal{E}}$ has the same dimension as $\hat{\mathbf{p}}_{\mathcal{V}_i}^{\mathcal{E}}$.
Both $\hat{\mathbf{p}}_{\mathcal{H}}^{\mathcal{E}}$ and $\hat{\mathbf{p}}_{\mathcal{H}}^{{\mathcal{V}_i}} $ have the same size as ${\mathbf{p}}_{\mathcal{H}}$.
As a result, the associated direct path wave propagation from the equivalent source plane to the hologram plane is
\begin{equation}
\begin{aligned}
    \hat{\mathbf{p}}_{\mathcal{E}} &= \Theta_1 (\hat{\mathbf{v}}_{\mathcal{E}}, \mathbf{s}', \omega), \\
\hat{\mathbf{p}}_{\mathcal{H}}^{\mathcal{E}} &=  \Theta_2 (\hat{\mathbf{p}}_{\mathcal{E}},\hat{\mathbf{v}}_{\mathcal{E}}, \mathbf{s}', \mathbf{r}, \omega).
\end{aligned}
\label{eq:direct_path}
\end{equation}
The paths involving virtual planes, utilized as an intermediate step (additional path), are
\begin{equation}
\begin{aligned}
     \hat{\mathbf{p}}_{\mathcal{V}_i}^{\mathcal{E}} &=  \Theta_2 (\hat{\mathbf{p}}_{\mathcal{E}},\hat{\mathbf{v}}_{\mathcal{E}}, \mathbf{s}', \mathbf{s}''_i, \omega),
     \\   \hat{\mathbf{v}}_{\mathcal{V}_i}^{\mathcal{E}} &=  \Theta_3 (\hat{\mathbf{p}}_{\mathcal{E}},\hat{\mathbf{v}}_{\mathcal{E}}, \mathbf{s}', \mathbf{s}''_i, \omega),
\\ \hat{\mathbf{p}}_{\mathcal{H}}^{\mathcal{V}_i} &=  \Theta_2 (\hat{\mathbf{p}}_{\mathcal{V}_i}^{\mathcal{E}},\hat{\mathbf{v}}_{\mathcal{V}_i}^{\mathcal{E}}, \mathbf{s}''_i, \mathbf{r}, \omega).
\end{aligned}
\label{eq:intermediate_path}
\end{equation}

Then the optimization problem consists in finding the velocity of the equivalent sources according to 
\begin{equation}
\begin{aligned}
   \hat{\mathbf{v}}_\mathcal{E} = \arg\min_{\mathbf{v}_\mathcal{E}} \bigg( \| {\mathbf{p}}_\mathcal{H} - \hat{\mathbf{p}}_\mathcal{H}^\mathcal{E}\|_a^a + \sum_{i=1}^{N_v}
    \| \mathbf{p}_\mathcal{H} - \hat{\mathbf{p}}_\mathcal{H}^{\mathcal{V}_i}\|_a^a
     + \\
    \lambda \| \hat{\mathbf{v}}_\mathcal{E} \|_b^b \bigg),
\end{aligned}
    \label{eq:pinn-fd-sesm-formulation}
\end{equation}
% where $\mathcal{V}_i$ denote the $i$-th virtual plane with $N_3$ spatial grids at $\mathbf{s}''_i$, and $i \in [1, N_v]$. The subscript represents the corresponding field,
% while the superscript indicates the originating field. 
% $\hat{\mathbf{v}}_\mathcal{E} = [\hat{v}(\mathbf{s}_1',\omega), ..., \hat{v}(\mathbf{s}_{N_1}',\omega)]^T \in \mathbb{C}^{N_1}$ is the reconstructed equivalent sources velocity, and both $\hat{\mathbf{p}}_\mathcal{H}^\mathcal{E}$ and $\hat{\mathbf{p}}_\mathcal{H}^{\mathcal{V}_i}$ have the same size as $\mathbf{p}_\mathcal{H}$.
It is important to highlight that this optimization formulation differs fundamentally from C-ESM, as it does not involve additional paths, as seen in \eqref{eq:cesm-formulation}. Additionally, this formulation is distinct from the approach in \cite{luancomplex, olivieri2021pinn}, where only the path between the actual source and the hologram is considered.

% Note that this discretized KH integral approach is similar to the methods in \cite{luancomplex, olivieri2021pinn}, however, only $\Theta_2$ \eqref{eq:ds-ph} was considered in those works.

% \textcolor{red}{check all notation and vectors form! $p_{\mathcal{H}} \in \mathbb{C}^M$ and $v_{\mathcal{E}} \in \mathbb{C}^{N_1}$.}
% The function $\Theta_1, \Theta_2$ and $\Theta_3$ represent the discrete KH integral equations given by \eqref{eq:ds-ps}, \eqref{eq:ds-ph} and \eqref{eq:ds-vh}, respectively, as the derivations in Appendix \ref{sec:appendix2} for planar coordinates.

% This is a patch-type Nearfield Acoustical Holography (NAH) method, where a small, planar microphone array is used to cover only a portion of the source surface. In the context of NAH, ``patch" refers to the fact that complete measurements over the entire source surface are not necessary.

%% FABIO: HERE
\subsection{Optimization via PINN}

% 4. optimization solver: PINN (sparse: both l1) with adam optimizer

A Physics-Informed Neural Network (PINN) is employed to solve the optimization problem in \eqref{eq:pinn-fd-sesm-formulation}, establishing a mapping from the input pressure field on the hologram plane $\mathbf{p}_{\mathcal{H}}$ to the output velocity field on the equivalent source plane $\hat{\mathbf{v}}_{\mathcal{E}}$, as
% Then the NAH problem formulation in PINN-FD-SESM is 
\begin{equation}
        \mathbf{p}_{\mathcal{H}}  := \Gamma (\boldsymbol{\gamma})  \hat{\mathbf{v}}_{\mathcal{E}}, \quad 
\hat{\mathbf{v}}_{\mathcal{E}}  = \Gamma (\boldsymbol{\gamma})^{-1}  \mathbf{p}_{\mathcal{H}},
\end{equation}
where $\Gamma (\boldsymbol{\gamma})$ is a discrete estimator that approximates $\mathbf{p}_{\mathcal{H}}$ given $\hat{\mathbf{v}}_{\mathcal{E}}$ by the PINN and $\boldsymbol{\gamma}$ are the trainable parameters of the NN. 
% $\Gamma$ serves a similar role to the Green's function in conventional C-ESM, while $\hat{v}(\mathbf{s},\omega)$ is analogous to $\hat{q}$.

\begin{figure*}
\vspace{-2mm}
    \centering
\includegraphics[width=0.9\linewidth]{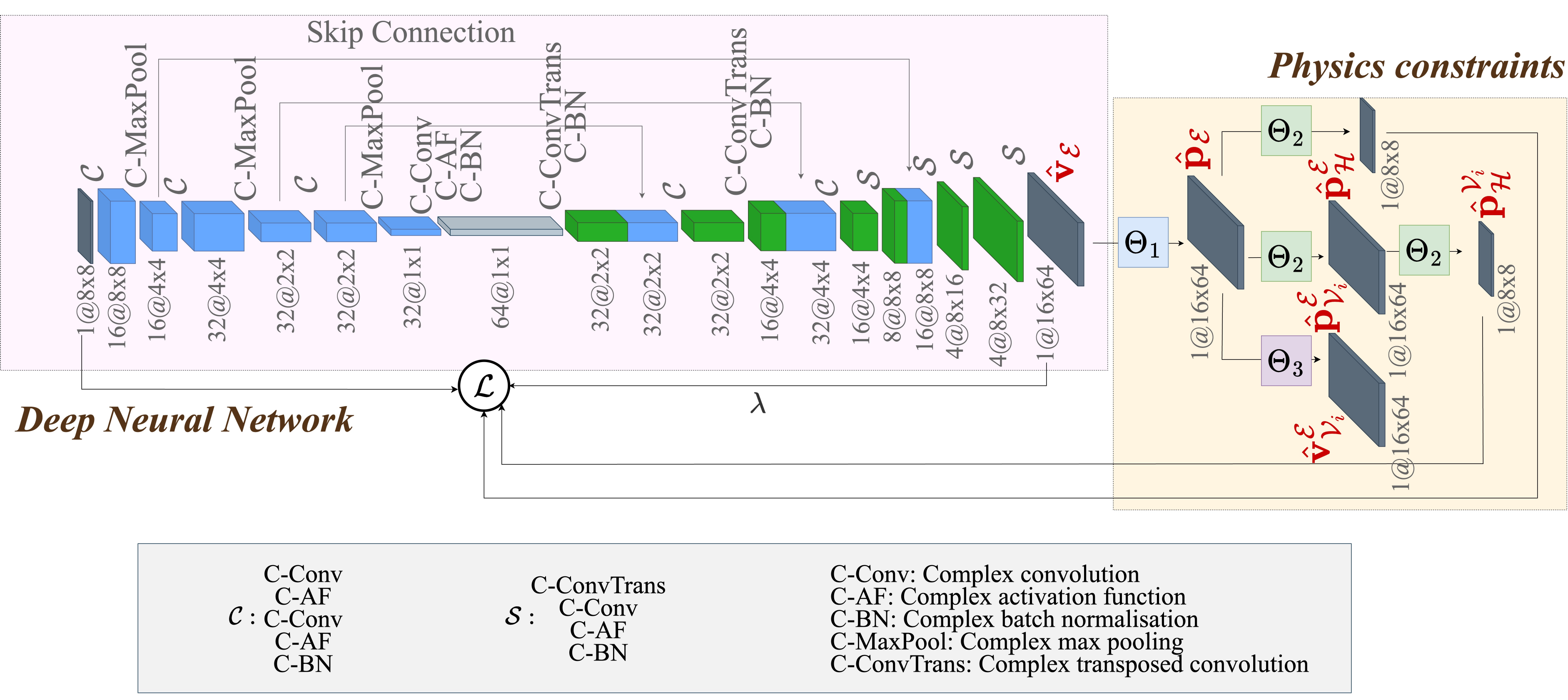}
    \caption{The architecture of PINN-SFD. $p_\mathcal{H}$ is fed in the encoder, through a bottleneck, followed by a decoder with output $\hat{v}_\mathcal{E}$, respectively. The network outputs then are fed into the following KH integral $\Theta_1$ \eqref{eq:ds-ps}, $\Theta_2$ \eqref{eq:ds-ph}, and  $\Theta_3$ \eqref{eq:ds-vh} to get the reconstruction fields, with known frequency and plane coordinates. The loss function consist the direct and additional paths pressure losses and a regularization term. The layer sizes and the operations used are indicated in the bottom box. The normalizations of the $\mathbf{p}_\mathcal{H}$ and $\hat{\mathbf{v}}_\mathcal{E}$ are not shown.}
    \label{fig:CESM_network}
    % \vspace{-2mm}
\end{figure*}

The NN architecture is a modified complex-valued \cite{hirose2012complex, trabelsi2017deep}  CNN, adapted from the model previously presented in \cite{luancomplex}.
The backbone of the model is a modified U-Net \cite{ronneberger2015u}, comprising an encoder, a bottleneck, and a decoder. Refer to Fig.~\ref{fig:CESM_network} for a detailed illustration of the NN architecture.
Additionally, complex activation function, convolution, transposed convolution, max pooling, and batch normalization are used to implement the complex-valued NN \cite{trabelsi2017deep}.
The complex activation function Cardioid, as proposed in \cite{virtue2017better}, is utilized in this work, with previous comparisons demonstrating its effectiveness in \cite{luancomplex}.  
Now, all of the constraints that are described in  \eqref{eq:pinn-fd-sesm-formulation}, with both the direct path \eqref{eq:direct_path} and additional path \eqref{eq:intermediate_path} propagations are incorporated into the NN loss functions, as
\begin{equation}\label{eq: loss}
\begin{aligned}
    \mathcal{L} = \frac{1}{M}\Big (\| \mathbf{p}_{\mathcal{H}}-\hat{\mathbf{p}}_{\mathcal{H}}^{\mathcal{E}} \|_1 
    + 
\sum_{i=1}^{N_v} \| \mathbf{p}_{\mathcal{H}}-\hat{\mathbf{p}}_{\mathcal{H}}^{\mathcal{V}i} \|_1 \Big )
+
    \lambda \| \hat{\mathbf{v}}_{\mathcal{E}} \|_1,
    \end{aligned}
\end{equation}
where  the parameter $\lambda>0$ controls the relative importance between the data fitting terms and the regularization term. Refer to Fig.~\ref{fig:CESM_network} for a block diagram illustrating the presentation of the physics loss. 
The Mean Absolute Error (MAE) is utilized rather than mean squared error, for the data fitting term of both direct and indirect paths, based on the experimental results. It indicates a preference for the $l_1$-norm than the $l_2$-norm. Notice that in conventional ESM, a $l_2 $-norm is used for the data fitting term in \eqref{eq:cesm-formulation}  ($a=2$) \cite{sarkissian2005method}. 
The sparsity is enforced by applying the $l_1$-norm of the penalty term on $\hat{\mathbf{v}}_{\mathcal{E}}$ as explained in Section \ref{sec: esm}, \eqref{eq:cesm-formulation},
which makes only a few equivalent sources active.
% Although several methods that combine $l_1$-norm and $l_2$-norm regularization have been proposed for the penalty term \cite{rahimi2019scale, pham2017noise, wang2020accelerated, huang2020ratio}, we do not introduce these innovative regularization technique in this paper; this could be an area for future enhancement.
To obtain the solution  $\hat{\mathbf{v}}_{\mathcal{E}}$, a gradient descent routine is applied during NN training using the Adaptive Moment Estimation (Adam) optimizer \cite{kingma2014adam}.

Once $\hat{\mathbf{v}}_{\mathcal{E}}$ is determined, the velocity at the actual source surface can then be reconstructed by
\begin{equation}
        \hat{\mathbf{v}}_{\mathcal{S}} = \Theta_3 (\hat{\mathbf{p}}_{\mathcal{E}},\hat{\mathbf{v}}_{\mathcal{E}}, \mathbf{s}', \mathbf{s}, \omega),
\end{equation}
with $\hat{\mathbf{p}}_{\mathcal{E}}$ can be computed by the first equation in \eqref{eq:direct_path}.

\subsubsection*{Remarks}

From the perspective of PINNs, the proposed PINN-SFD framework utilizes NNs as a physics-informed regularization agent, aligning with traditional PINNs \cite{raissi2019physics}, but it fundamentally differs from the previous approach \cite{luancomplex, olivieri2021pinn}. 
Indeed the architecture proposed in \cite{luancomplex} is a hybrid of PGNN and PINN, while PINN-SFD is a pure PINN.
Specifically, the PINN-SFD operates in a single-instant, self-supervised manner: \textit{single-instant} indicates that the optimization problem is based on a single sample without necessitating a large training dataset, while \textit{self-supervised} signifies that the optimization is driven by the physics-informed regularization agent, eliminating the need for ground truth data.

\section{Validation of proposed method}\label{sec:validation}
\subsection{Implementation}
% \begin{itemize}
%     \item dataset
%     \item GPU
%     \item training setup ...
%     \item hyperparameters
% \end{itemize}
\subsubsection{Models}\label{sec:models}
% We assume that the spatial grids on the real source surface align with those of the equivalent sources.
The NAH configuration is as follows:
Plane $\mathcal{S}$ is at $z_\mathcal{S} = 0$, $\mathcal{H}$ is at $z_\mathcal{H} = \SI{3.12}{\centi\meter}$, and $\mathcal{E}$ is at $z_\mathcal{E} = \SI{-5.0}{\centi\meter}$. 
Three VPs $\mathcal{V}_i, i \in [1,3]$ are positioned near the source plane, at $z_{\mathcal{V}_1}=z_\mathcal{S} = 0$, $z_{\mathcal{V}_2} = $ $\SI{-0.10}{\centi\meter}$, and $z_{\mathcal{V}_3} = $ $\SI{0.10}{\centi\meter}$. The grids on the planes $\mathcal{E}$ and $\mathcal{V}_i$ have the same number of points on the one on $\mathcal{S}$, as $N_1 = N_2= N_3 = 16 \times 64$. The number of points on the grid on $\mathcal{H}$ is $M = 8 \times 8$. Note that this is an up-sampling procedure, as $N_2 > M$.

To evaluate the effectiveness of the VPs, we employ two models: one that incorporates VPs, with $N_v =3$, and another that does not, i.e. $N_v = 0$. A fixed regularization parameter of $\lambda = 1 \times 10^{-6}$ is applied in both cases. 
%This value has been found as the one that guarantees the best accuracy in a preliminary experiment. 
% The experimental results indicate that the prediction remains largely unaffected whether a fixed or variable regularization parameter is employed.
\textit{Pytorch} \cite{paszke2019pytorch} is used for the NN implementation and the library \textit{complexPyTorch} \cite{matthes2021learning} is used to implement complex-valued NNs in this study. 
Moreover, the conventional C-ESM \cite{fernandez2017sparse} is also implemented using the MATLAB toolbox CVX \cite{grant2014cvx} for comparison. Five regularization parameters are applied, evenly spaced within the range of $[0.001, 0.1]$ for violin top plate dataset and $[0.005, 0.1]$ for rectangular plate dataset. The final result is selected based on the best reconstruction of the hologram pressure field with the minimum MAE loss. 

\subsubsection*{Remark} 
The experimental results with PINN-SFD indicate that the predictions remain largely consistent, whether a fixed or variable regularization parameter is employed, which is why we opted for a fixed value. However, for conventional C-ESM, the prediction outcomes are significantly influenced by the regularization parameter value \cite{fernandez2017sparse}. This suggests that PINN-SFD is relatively less sensitive to the choice of regularization parameter compared to C-ESM, which is advantageous as it reduces the need for manual tuning of this parameter.

\subsubsection{Dataset}
In this study, the NAH dataset used, as described in \cite{olivieri2021pinn}, consists of objects such as violin top plates and rectangular plates with various boundary conditions (clamped, free, and simply supported). The eigenfrequency for the dataset generation is limited in $[0, 2000]$ $\si{\hertz}$.
% Unlike the typical study on PINNs \cite{raissi2019physics} that evaluate a single sample due to the one-shot procedure, we assess the robustness of the proposed method across various cases. 
To this end, we selected ten violin plates, each covering all modes within the restricted frequency band from the complete violin dataset, yielding a total of 471 samples. It is important to note that the highest mode number in the limited frequency band may vary across different plates. The highest mode number observed among the the considered ten violin plates in the dataset reaches 53.
For the rectangular plate dataset, unlike the violin case where samples were chosen from a fixed number of plates, we used a selection criterion based on mode numbers. Specifically, ten samples were selected for each mode, with the following ranges: 1–14 for clamped boundary conditions, 1–30 for free boundary conditions, and 1–20 for simply supported boundary conditions. However, for the highest modes, there may be slightly fewer than ten samples.
This selection process results in 139 samples for clamped, 299 for free, and 183 for simply supported boundary conditions. A summary of the utilized dataset is shown in Table~\ref{table:dataset}.
Additionally, it is noteworthy that the training runs are independent for each sample, adhering to a single-instant procedure.

\begin{table}[h]
{\fontsize{10}{12}\selectfont
    \centering
    \caption{Summary of selected samples from violin top plate and rectangular plate datasets. ``Rec." refers to rectangular plates, ``Cl." to clamped, ``Fr." to free, and ``Si." to simply supported.}
    \begin{tabular}{l l l l}
        \hline
        \textbf{Dataset} & \textbf{BC}    & \textbf{\# of Modes} & \textbf{\# of Samples}   \\ \hline
        Violin  & -    & All modes (1-53) & 471 (10 plates)   \\  \hline
        Rec.  & Cl.   & 1–14  & 139       \\ 
      & Fr.  & 1–30  & 299  \\ 
      & Si.   & 1–20   & 183                          \\ \hline
    \end{tabular}
    \label{table:dataset}
    }
\end{table}

The same data augmentation and additive noise procedure is used according to \cite{olivieri2021pinn}. 
The additive noise applied to each pressure item in the datasets has a signal-to-noise ratio (SNR) selected from a uniform distribution in the interval [10, 60] \si{\decibel}.

In order to facilitate easier training of the NN, a normalization-like procedure is adopted, using the normalization factors  $\alpha$ and $\beta$, defined as $ \beta = \alpha \max{|\mathbf{p}_\mathcal{H}|}$.
Consequently, the NN input is normalized as $
\mathbf{p}_\mathcal{H} / \beta$, while the NN output is denormalized to $ \beta \hat{\mathbf{v}}_\mathcal{E}$.

% \subsubsection*{Remark 2}
% In PINN-SFD, the equivalent source pressure is fully determined by the source velocity through the discretized KH integral $\Theta_1$ \eqref{eq:ds-ps}. 
% By contrast, in the previous CV-KHCNN framework \cite{luancomplex}, the actual source pressure and velocity are indirectly connected, as they are the two separate outputs of the NN with distinct decoders. In CV-KHCNN, constraints then link these two outputs with the hologram pressure via $\Theta_2$ \eqref{eq:ds-ph} within the loss function. This indirect relationship, along with the normalization of inputs and outputs in CV-KHCNN (where a consistent normalization factor cannot be determined, since all samples are normalized to $[-1, 1]$ when training a large dataset simultaneously), introduces additional flexibility in the solution space. This flexibility can lead to uncertainty in the loss function’s scaling, which impacts physical accuracy. However, PINN-SFD maintains physical accuracy in wave propagation by enforcing the physics law.

\subsubsection{Hyperparameters}
The normalization factor is $\alpha = 1\%$.
The Adam optimizer with an initial learning rate of 0.01, which is reduced by a factor of 0.1 after 200 epochs without improvement, until it reaches 0.001. Early stopping is implemented to prevent overfitting, halting the training after 50 epochs with no improvement in the loss function.
The initialization of the NN parameters $\boldsymbol{\gamma}$ all follow the default settings. 

\subsection{Metrics}
 The performance of the proposed network is assessed by two metrics: the Normalized Mean Square Error ($\operatorname{NMSE}$) (defined in $\si{\decibel}$) and the Normalized Cross Correlation ($\operatorname{NCC}$). They are expressed by
\begin{equation}
    \operatorname{NMSE}(\hat{\mathbf{x}},\mathbf{x}) = 10\mathrm{log}_{10}\left ( \frac{\mathbf{e}^H \cdot \mathbf{e}}{\mathbf{x}^H \cdot \mathbf{x}} \right ),
\end{equation}
and
\begin{equation}
    \operatorname{NCC}(\hat{\mathbf{x}},\mathbf{x}) = \frac{\hat{\mathbf{x}}^H \cdot \mathbf{x}  }{\|\hat{\mathbf{x}} \|_2 \cdot \| \mathbf{x}\|_2},
\end{equation}
where $\mathbf{x}$ are the ground truth data, $\hat{\mathbf{x}}$ are the predicted data, and $\mathbf{e} = \hat{\mathbf{x}} - \mathbf{x}$ denote the error. Additionally, the metrics are computed with a column-vector representation of the data, and $\operatorname{NCC}$ reaches 1 when the two quantities match perfectly.
Note that both the metrics are for complex numbers and $^H$ is the Hermitian transpose operator.

It is worth to mention that since the violin plate has irregular shape, binary mask is adopted to select the points of the mesh grid belonging to the target surface when evaluating the surface plane of the violin top plates, as done in \cite{olivieri2021pinn}.

\subsection{Results}
\begin{figure*}
    \centering
    \subfloat[\parbox{2cm}{\centering \scriptsize C-ESM \\ \scriptsize{$\operatorname{NMSE}$: -4.04} \\ \scriptsize{$\operatorname{NCC}$: 79.27\%}}]{
        \includegraphics[scale=0.5]{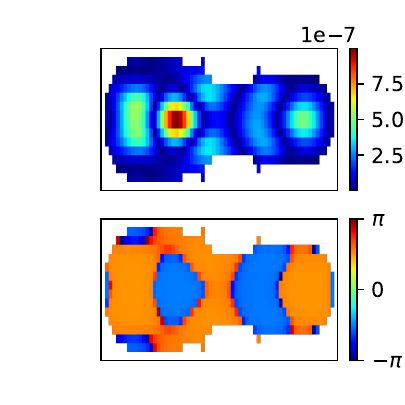}
    }
    \subfloat[\parbox{3cm}{\centering  \scriptsize PINN-SFD ($N_v=0$) \\ \scriptsize{$\operatorname{NMSE}$: -3.70} \\ \scriptsize{$\operatorname{NCC}$: 84.47\%}}]{
        \includegraphics[scale=0.5]{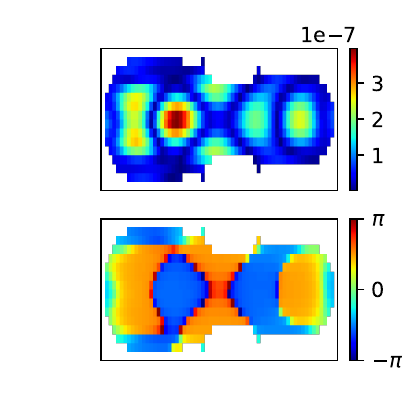}
    }
    \subfloat[\parbox{3cm}{\centering \scriptsize PINN-SFD ($N_v = 3$) \\ \scriptsize{$\operatorname{NMSE}$: -5.60} \\ \scriptsize{$\operatorname{NCC}$: 86.41\%}}]{
        \includegraphics[scale=0.5]{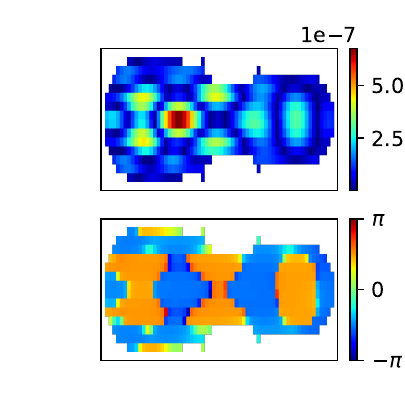}
    }
    \subfloat[\scriptsize Ground truth]{
        \includegraphics[scale=0.5]{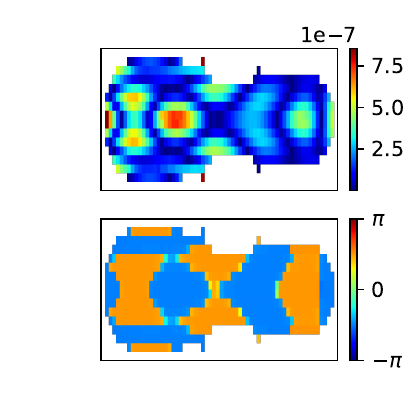}
    }
    \caption{An example of $\textbf{V}_\mathcal{S}$ for a violin top plate of mode 23, at 985.30 \si{\hertz}. Top: Magnitude. Bottom: Phase.}
    \label{fig:sample_violin 23}
\end{figure*}

% \subsubsection{Examples}
The source velocity fields reconstructed by C-ESM, PINN-SFD ($N_v=0$ and $N_v=3$) along with the ground truth, for two samples are shown in
Fig.~\ref{fig: violin_NCC} and \ref{fig: rec_free_NCC}.
In these two cases, both PINN-SFD, $N_v=0$ and $N_v=3$ outperform C-ESM, with the difference being more evident when examining the phase.
By comparing the results of PINN-SFD, $N_v=0$  and $N_v=3$ shown in Fig.~\ref{fig:sample_violin 23}, it is evident that the VPs improves the representation of fine details, such as intricate structures, textures, and edges, for high-complexity patterns.
Conversely, for rectangular plates, as visible in Fig.~\ref{fig: rec_free_NCC}, both PINN-SFD, $N_v=0$ and $N_v=3$  can ensure similarity in the basic shape, unlike C-ESM.

\begin{figure*}
    \centering
    \subfloat[\parbox{2cm}{\centering \scriptsize C-ESM\\  \scriptsize{$\operatorname{NMSE}$: 0.16} \\  \scriptsize{$\operatorname{NCC}$: 58.50\%}}]{
        \includegraphics[scale=0.5]{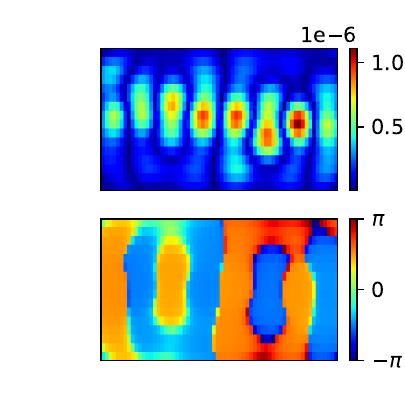}
    }
    \subfloat[\parbox{3cm}{\centering \scriptsize PINN-SFD ($N_v =0$)\\  \scriptsize{$\operatorname{NMSE}$: -3.98} \\  \scriptsize{$\operatorname{NCC}$: 86.84\%}}]{
        \includegraphics[scale=0.5]{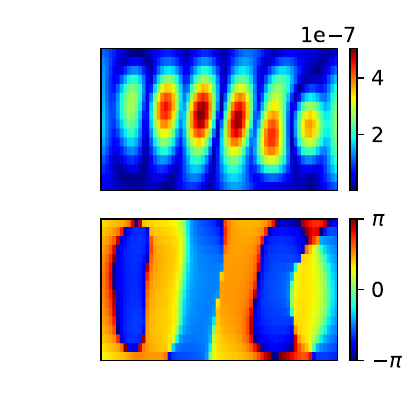}
    }
    \subfloat [\parbox{3cm}{\centering \scriptsize  PINN-SFD ($N_v =3$) \\ \scriptsize{$\operatorname{NMSE}$: -6.28} \\ \scriptsize{$\operatorname{NCC}$: 88.26\%}}]{
        \includegraphics[scale=0.5]{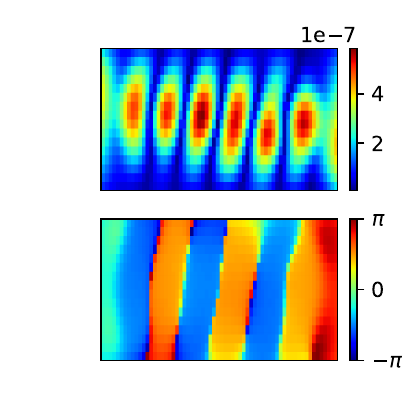}
    }
    \subfloat[\scriptsize Ground truth]{
        \includegraphics[scale=0.5]{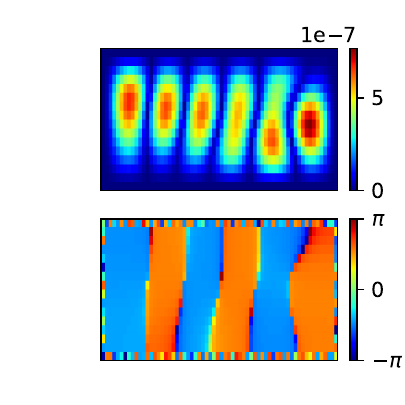}
    }
    \caption{An example of $\textbf{V}_\mathcal{S}$ for a clamped rectangular plate of mode 14, at 1964.02 \si{\hertz}. Top: Magnitude. Bottom: Phase.}
    \label{fig:sample_clamped14}
\end{figure*}

% \subsubsection{Overall results}
\begin{table*}
{\fontsize{10}{12}\selectfont
\begin{center}
\caption{Mean $\operatorname{NMSE}$ [$\si{\decibel}$] and $\operatorname{NCC}$ values of $\hat{\textbf{v}}_\mathcal{S}$ for different models and datasets.}
\begin{tabular}{p{3cm}*8{p{1.3cm}}}
\hline
\textbf{} & \multicolumn{2}{c}{\textbf{Violin}} & \multicolumn{2}{c}{\textbf{Rec. (Cl.)}} & \multicolumn{2}{c}{\textbf{Rec. (Fr.)}} & \multicolumn{2}{c}{\textbf{Rec. (Si.)}} \\
\cline{2-9} 
\textbf{} & \textit{NMSE}  &  \textit{NCC} & \textit{NMSE} &  \textit{NCC} & \textit{NMSE} &  \textit{NCC} & \textit{NMSE} &  \textit{NCC}  \\
\hline
C-ESM & -2.54 & 65.46\% & -2.63 & 73.51\% &-1.63 & 59.61\% & -4.17 & 76.72\%\\
PINN-SFD ($N_v=0$) &  -2.01 & 67.69\% & -4.27 & \textbf{89.93\%}& -1.84 & \textbf{68.53\%} & -4.50 & \textbf{90.00\%}\\
PINN-SFD ($N_v=3$) & \textbf{-2.79} & \textbf{70.83\%} & \textbf{-5.07} & 84.04\% & \textbf{-2.35} & 68.23\% & \textbf{-5.05} & 84.07\% \\
\hline
\multicolumn{7}{l}{$^*$Values marked in bold are the best performances.} & \multicolumn{2}{l}{} \\
\end{tabular}
\label{table: PINN-VP-CESM}
\end{center}
% \vspace{-3mm}
\vspace{-3mm}
} 
\end{table*}

The mean $\operatorname{NMSE}$ and $\operatorname{NCC}$ values of the reconstructed actual source velocity field for each dataset, obtained using C-ESM, PINN-SFD ($N_v=0$ and $N_v=3$), are presented in Table~\ref{table: PINN-VP-CESM}.
The results indicate that both PINN-SFD, $N_v=0$ and $N_v=3$ consistently demonstrate better performance compared to C-ESM in terms of $\operatorname{NMSE}$ and $\operatorname{NCC}$. This finding underscores the advantages of PINN optimization over traditional methods such as interior-point method in \cite{grant2014cvx}.

However, it is noteworthy that PINN-SPD, $N_v=3$ does not always outperform $N_v=0$, suggesting that the introduction of the VPs can have different effects on the optimization process. While the presence of the VPs tends to consistently improve $\operatorname{NMSE}$, it may result in a decrease in $\operatorname{NCC}$. This indicates that even though the error between the prediction and the ground truth is small, the pattern similarity may not be adequately enhanced.
% One possible hypothesis is that the additional constraints of VP are more effective in source reconstruction with more intricate patterns or textures. 

% It can be observed that both PINN-SESM and PINN-FD-SESM generally outperform C-ESM in terms of both NMSE and NCC, which highlights the efficiency of PINN optimization over traditional method (CVX). However, PINN-VP-CESM not always outperforms PINN-CESM, which means the additional virtual planes sometimes have a positive effect.

An intriguing aspect is the reduced sensitivity of PINN-SFD to regularization parameters, whereas conventional methods heavily rely on the selection of regularization parameters \cite{fernandez2017sparse}, as remarked in Sec~\ref{sec:models}. Overall, in terms of metrics, PINN-SFD exhibits smoother reconstruction performance. Despite PINN-SFD employing a constant regularization parameter, even when C-ESM selects the best hologram pressure reconstruction among five regularization parameters, PINN-SFD's results remain more stable than C-ESM. PINN-SFD overcomes the limitation of traditional methods that depend heavily on regularization parameters, offering a more robust approach.
Additionally, for a fair comparison of the three methods, the loss depends on hologram pressure, only. It can be observed that constraining hologram pressure reconstruction to select C-ESM regularization parameters is not particularly effective. The selection of regularization parameters in C-ESM remains a challenge. Typically, the selection of the regularization parameter in C-ESM involves an iterative solution of the problem along with prior knowledge of the relative noise level \cite{fernandez2017sparse, gerstoft2015multiple}. Furthermore, when selecting the regularization parameter for C-ESM in this study, we found that it is significantly related to the complexity of the source's vibrational pattern.

\subsubsection{Accuracy across modes}\label{sec: accur}
\begin{figure*}[!t]
\centering
\subfloat[]{\includegraphics[width=1\linewidth]
{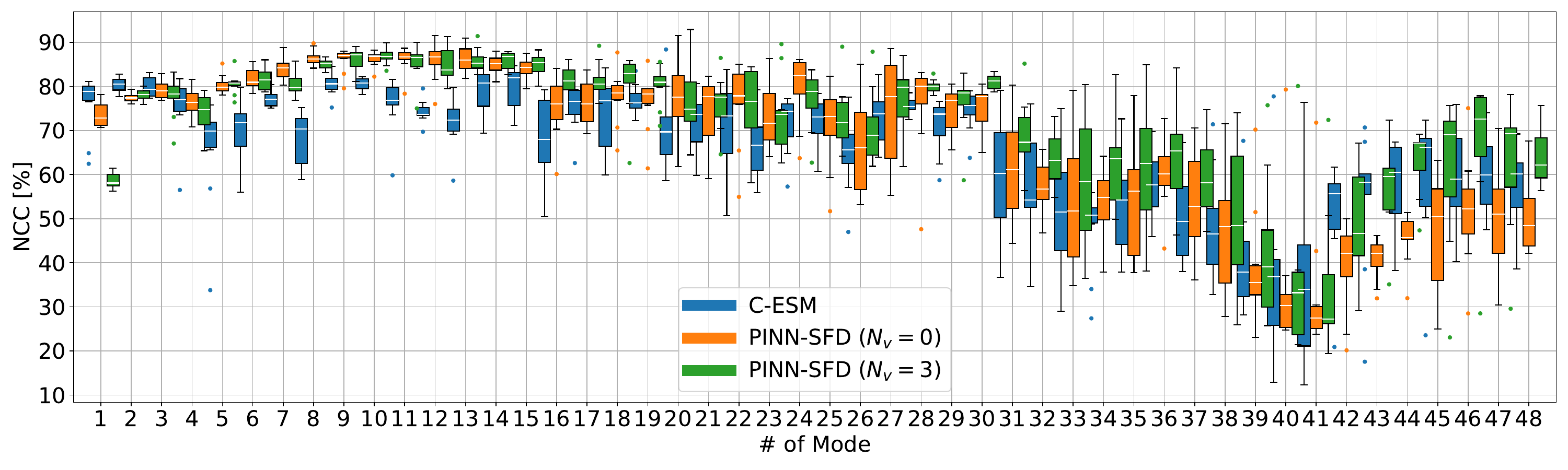}%
\label{fig: violin_NCC}}
% \hfil
\\
\subfloat[]{\includegraphics[width=1\linewidth]
{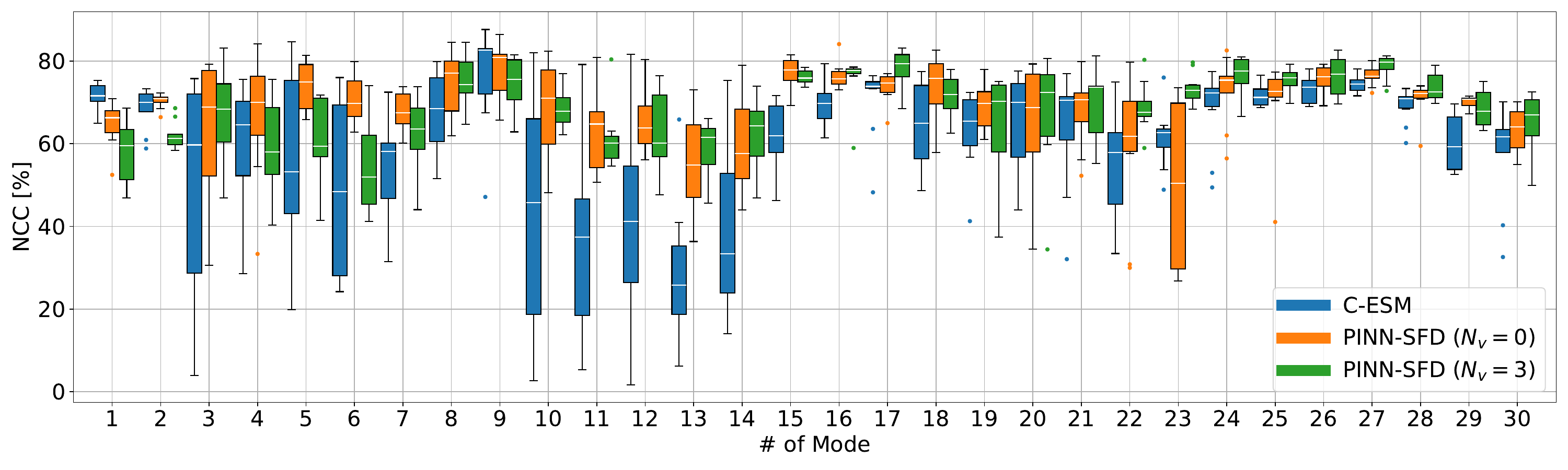}%
\label{fig: rec_free_NCC}}
\caption{Mean $\operatorname{NCC}$  of $\hat{\textbf{V}}_\mathcal{S}$ across the modes. (a) Violin. (b) Rectangular plate with free BCs.}
\label{fig: NCC}
\end{figure*}

To differentiate the behavior of various modes across different datasets, the $\operatorname{NCC}$  for the modes of the rectangular plate with free BCs and the violin dataset are presented in Fig.~\ref{fig: NCC}. We chose to present the results for the rectangular plate with free BCs instead of other BCs because it exhibits a wider variety of complex vibrational patterns, which represents a more challenging scenario.
Furthermore, due to the limited number of violin samples for modes beyond 48, they are not visualized in Fig.~\ref{fig: NCC}.

The results of both PINN-SFD, $N_v=0$ and $N_v=3$  exhibit smaller fluctuations compared to C-ESM, further demonstrating the PINN's enhanced stability.
% Also, it is evident that the source field reconstruction accuracy is related to the mode, indicating a close relationship to the frequency and vibrational pattern (typically, higher modes have more complex patterns). 
The trend of $\operatorname{NCC}$  for C-ESM, PINN-SFD ($N_v=0$ and $N_v=3$) of violin top plates is similar, with C-ESM generally exhibiting a downward shift overall.  However, a notable exception is for the 1st mode of both PINN-SFD, $N_v=0$ and $N_v=3$, which perform worse than C-ESM,  with $N_v=3$ with even poorer results. 
This may be due to overfitting of the PINN-SFD training.
The accuracy remains relatively high up to around the 30th mode, but for the higher modes, it decreases, reaching a low point around the 40th mode, before increasing again. 
However, the performance metrics of rectangular plates show lower accuracy in low modes compared to high modes. This trend shows consistently with different BCs. Once again, the 1st mode exhibits the same tendency as observed for the violin.
Before the 15th mode, C-ESM shows significant fluctuations and much lower accuracy compared to both PINN-SFD, $N_v=0$ and $N_v=3$.

This analysis suggests that the prediction accuracy is likely related to the complexity of the vibrational patterns. 
Indeed, violin top plates have more complex vibrational patterns than rectangular plates, due to their irregular shape and orthotropic wood properties. 
This may result in the violin dataset lacking the simple patterns found in the low-frequency modes of rectangular plates.
Moreover, since the considered rectangular plate dataset does not include very high frequency modes, it might not reach the limit of overly complex patterns.

Related to the complexity of the vibrational patterns, it can be inferred that  PINN-SFD for $N_v=0$ and $N_v=3$ has higher reconstruction ability for complex vibrational patterns, such as violin top plate modes 2–30 and free rectangular plate modes 15–30. In this case, the VPs consistently show a positive effect that aids in improving predictions.
For simple patterns in rectangular plate, such as modes 2–14,  PINN-SFD, $N_v=0$ and $N_v=3$ significantly outperform C-ESM, while the presence of VPs does not always bring a notable improvement.
PINN-SFD for $N_v=3$ generally outperforms C-ESM for highly complex patterns in the violin dataset, such as modes 31–48. However, in the absence of the VPs, PINN-SFD with $N_v=0$ performs overall worse than C-ESM.

% Combining the NCC trends and the characteristic of the vibrational objects, we can infer the classification of pattern complexity of the evaluated dataset, shown in table \ref{table: complexity}. 
% \begin{itemize}
%     \item \textit{overly simple}: 1st mode for violin and rectangular plate with free BCs.
%     \item \textit{simple}: 2nd to 14th modes for rectangular plate with free BCs.
%     \item \textit{complex}: 2nd to 30th modes for violin and 15th to 30th modes for rectangular plate with free BCs.
%     \item \textit{overly complex}: 31st to 48th modes for violin.
% \end{itemize}
% \begin{table}
% {\fontsize{10}{12}\selectfont
% \begin{center}
% \caption{Classification of pattern complexity based on mode numbers for the evaluated dataset of the violin top plate and rectangular plate with free BCs.}
% \begin{tabular}{p{1cm}*4{p{1.2cm}}}
% \hline
% \textbf{} & {\textit{overly simple}} & {\textit{simple}} & {\textit{complex}} & {\textit{overly complex}} \\
% \hline
% \textbf{Violin} & 1 & - & 2-30 & 51-48\\
% \textbf{Rec.(fr.)} & 1 & 2-14 & 15-30 & -\\
% \hline
% \end{tabular}
%  \label{table: complexity}
% \end{center}
% % \vspace{-3mm}
% } 
% \end{table}

This result is consistent with the discussion in \cite{fernandez2017sparse}. The sparse solution of the equivalent source demonstrates higher performance for the non-redundant representation of the observed data. This underscores the inherent limitation of sparsity representation in compressive sensing, regarding the regularization term in \eqref{eq: loss}.

%%%%%%%%%%%%%%%%%%New version based on NMSE

\subsubsection{Tracking Actual Source Velocity During Training}
We recognize that NAH is a highly ill-posed inverse problem \cite{williams2001regularization}, where the optimization is contingent upon minimizing the loss associated with the hologram pressure.
% We know that NAH is an inverse problem, and the optimization depends on the loss of the hologram pressure.
Nevertheless, better hologram pressure reconstruction does not necessarily mean better actual source velocity reconstruction. 
% This issue is often overlooked in conventional C-ESM. 
Most algorithm comparisons in NAH focus only on the accuracy of the pressure field reconstruction \cite{hald2018comparison, hald2020comparison}, while there is a lack of consideration for the reconstruction of the actual source velocity. Obtaining ground truth actual source surface velocity patterns of vibrating objects through experiments is not easy and requires laser Doppler vibrometer or other instruments.  By leveraging simulation data, we can evaluate the surface velocity more effectively and provide a more comprehensive assessment of the algorithms' performance.

\begin{figure}[!t]
\centering
\subfloat[]{\includegraphics[width=1\linewidth]
{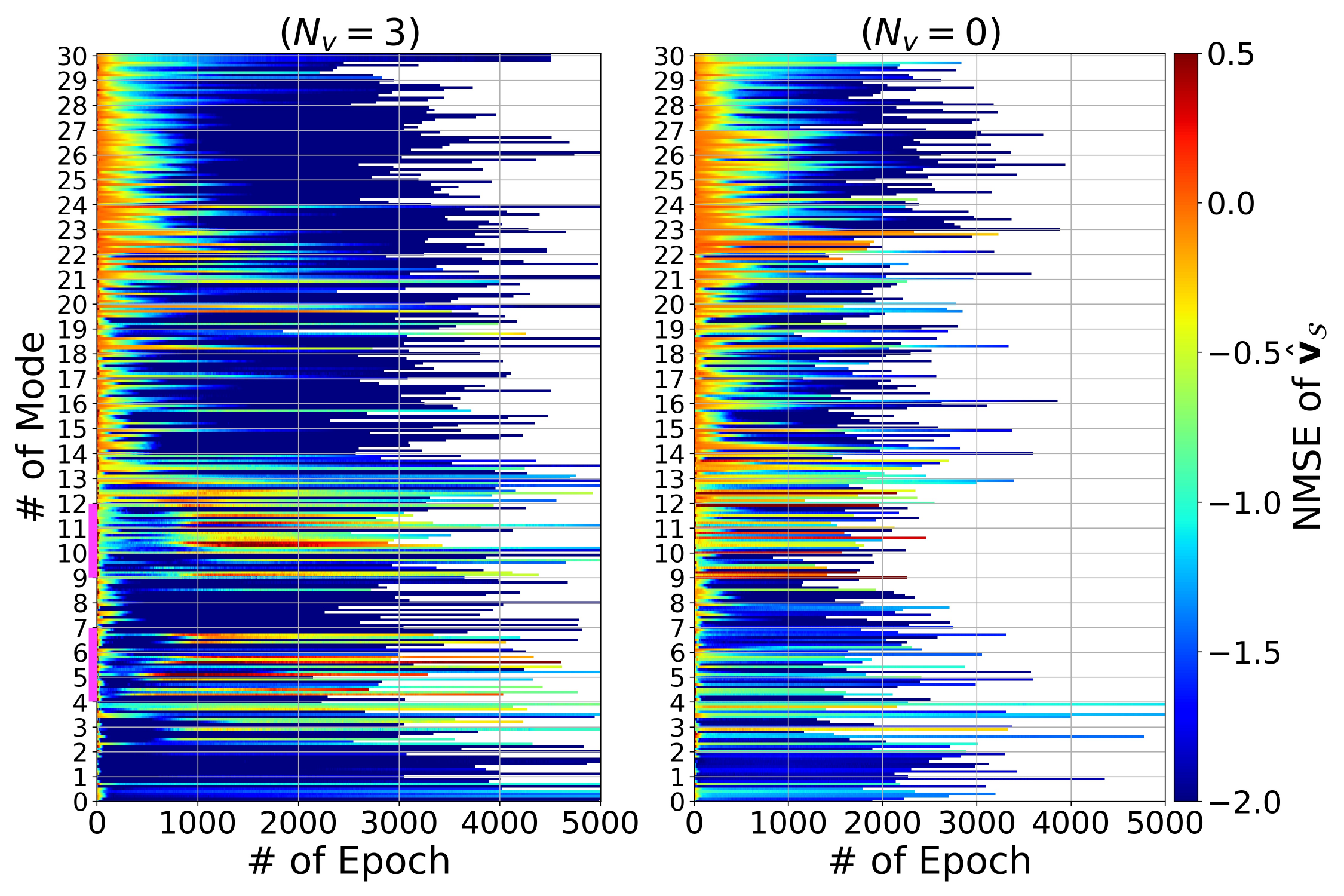}%
\label{fig:NMSE_epochs_rec_free}}
% \hfil
\\
\subfloat[]{\includegraphics[width=1\linewidth]
{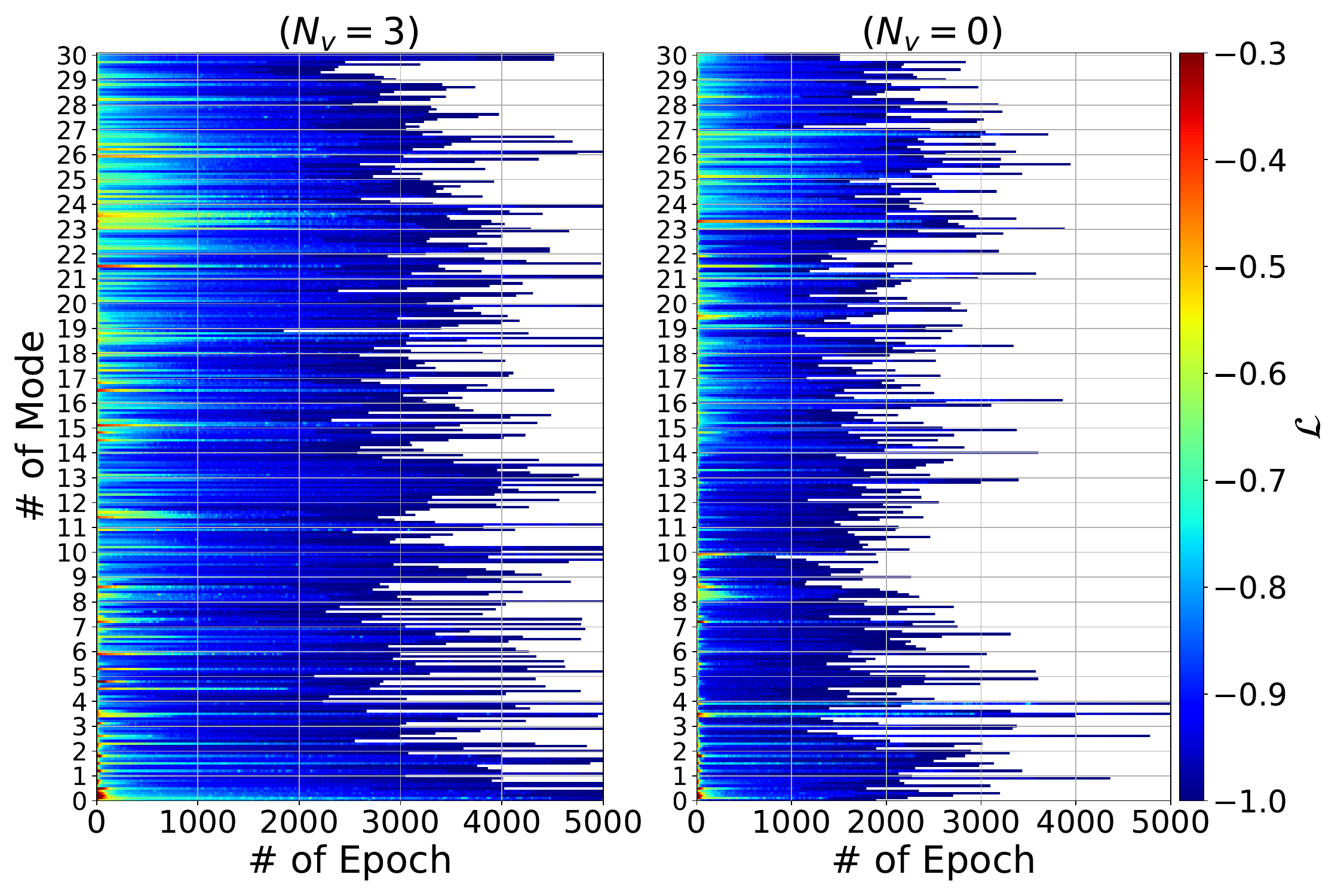}%
\label{fig:LOSS_epochs_rec_free}}
\caption{The evolution of the $\operatorname{NMSE}$ for $\hat{\mathbf{v}}_\mathcal{S}$ (top) and the training loss $\mathcal{L}$ (bottom) of free rectangular plates during training procedure for PINN-SFD, $N_v = 3$ (left) and $N_v = 0$ (right).}
\label{fig:epochs_rec_free}
\end{figure}

In order to understand how surface velocity evolves during the PINN-SFD optimization process, Fig.~\ref{fig:epochs_rec_free} presents the evolution of the $\operatorname{NMSE}$ of actual source velocity and the training loss throughout the training process of free rectangular plates for PINN-SFD for $N_v=0$ and $N_v=3$. 
It is noteworthy that the training loss defined in \eqref{eq: loss} is first normalized to the range $[0.1, 1.1]$ and then transformed using $\log_{10}$ for improved visualization.
% It is noteworthy that the figure demonstrates the correlation between the evolution of the source velocity $\operatorname{NMSE}$ and the training epochs during the training process.
The general trend of the $\operatorname{NMSE}$ in Fig.~\ref{fig:NMSE_epochs_rec_free} shows a decrease in both cases. However, in the mode ranges $[5, 7]$ and $[10, 12]$ for $N_v=3$, a different pattern emerges where the $\operatorname{NMSE}$ initially decreases to a certain point, then increases, and subsequently decreases again. We refer to this as the \textit{rebound effect}.
It is evident that around 1000 epochs, the velocity has already deviated from the previously found optimal value, and as it begins to decrease again, it may not reach the previously achieved lower value.
The rebound effect is particularly evident in the low-frequency modes and with an increased number of VPs. This may explain why, as mentioned in Section~\ref{sec: accur}, PINN-SFD with $N_v=3$ sometimes performs less effectively than $N_v=0$ at lower frequencies (as seen in Fig.~\ref{fig: rec_free_NCC}).
At high frequencies, the VPs indeed leads to faster convergence, as seen by comparing the color intensity of $N_v=0$ and $N_v=3$ at around 1000 epochs.
Meanwhile, the convergence of the training loss is ensured for both cases, as observed in Fig.~\ref{fig:LOSS_epochs_rec_free}.

The above discussion suggests that evaluating the reconstruction accuracy of hologram pressure alone is insufficient.
This analysis further indicates that the complexity of the vibration pattern directly affects the accuracy of the reconstruction. Integrating the complexity of the vibration pattern into the optimization algorithm is an issue that needs to be addressed. 
For PINN-SFD, when there are more VPs, implementing an effective early stopping strategy to mitigate the  rebound effect observed at low frequencies during training poses a significant challenge. It may be feasible to address this challenge by considering the complexity of the vibration pattern.

\section{Conclusion}\label{sec:conclusion}

In this study, we presented the PINN-SFD as a novel single-instant, self-supervised, physics-informed deep learning approach to solve the NAH problem. 
The wave propagation field is discretized into sparse regions, including the equivalent source plane and virtual planes, to solve the inverse optimization problem. 
Wave propagation is modeled using the discretized KH integral. By incorporating virtual planes, additional constraints are enforced near the actual sound source, improving the reconstruction process. 
Furthermore, optimization is carried out using PINNs,
where physics-based constraints are integrated into the loss
functions to account for both direct and indirect wave propagation paths.
Sparsity is promoted by introducing a regularization term for the equivalent source velocities in the loss functions during optimization.
% In the proposed PINN-SFD method, Field Discretization refers to the discretization of the wave propagation field through both the equivalent source plane and the virtual planes, while Sparse reflects the sparsity imposed both in the discretized field and the optimization process.
% Our method leverages the discretized KH integral for wave propagation model, setting it apart from C-ESM approaches that rely on wave-superposition. By integrating FD, we impose additional constraints near the actual sound source, enhancing the reconstruction process.
% In addtion, sparsity is imposed by adding a regularization term of the equivalent sources velocity in loss functions of the optimization process.

The validation of PINN-SFD on various rectangular plates and violin top plates revealed its better performance compared to C-ESM across a diverse range of vibrational modes. 
Moreover, PINN-SFD exhibits a reduced sensitivity to regularization parameters, which can be advantageous in practical applications.
Our findings highlight the intricate relationship between the complexity of the vibrational patterns and reconstruction accuracy. PINN-SFD excels in reconstructing detailed features of complex patterns. An interesting aspect is that VPs improve the reconstruction of fine details, especially in highly complex vibrational patterns.
However, when tracking the accuracy of the actual source velocity along training epochs, we observe that the rebound effect may occur in low-frequency modes when utilizing VPs.

In the future, it would be interesting to treat the number and positions of VPs as trainable parameters for optimization.
Tuning the NN architecture utilizing adaptive techniques for balancing loss functions may enhance the performance.
In addition, designing a metric to assess the vibration pattern and integrating it for adaptive hyperparameter selection holds promise for improving reconstruction accuracy and preventing overfitting.
It would also be beneficial to assess the proposed model using real measurement data. Furthermore, considering the Helmholtz equation instead of the KH integral as the governing equation offers an intriguing opportunity for further investigation.

% \textcolor{red}{Transfer learning? governing equation: Helmholtz equation?}

% \appendix%[...]
% \input{sections/appendix}
% or


%===========================
\section{}\label{sec:appendix1}

By substituting \eqref{eq:euler} into \eqref{eq:kh}, the pressure field generated by a vibrating structure is 

\begin{equation} \label{eq:kh2}
p(\mathbf{r}, \omega) = 
\begin{cases}
     \displaystyle \int_\mathcal{S} p(\mathbf{s}, \omega) \frac{\partial}{\partial \mathbf{n}} g_{\omega}(\mathbf{r},\mathbf{s}) d\mathcal{S} 
     - \\ j \omega \rho_0 \int_\mathcal{S} v_{n}(\mathbf{s},\omega) g_{\omega}(\mathbf{r},\mathbf{s}) d\mathcal{S}, 
     &\text{if $\mathbf{r}$ is outside $\mathcal{S}$} \\
    -2 j \omega \rho_0 \displaystyle \int_\mathcal{S} v_{n}(\mathbf{s},\omega) g_{\omega}(\mathbf{r},\mathbf{s}) d\mathcal{S}, &\text{if $\mathbf{r}$ is on $\mathcal{S}$}
 \end{cases}.
\end{equation}
Notice that when $\textbf{r}$ and $\textbf{s}$ are both on the surface $\mathcal{S}$, the first term of (\ref{eq:kh}) is zero and Green's function (\ref{eq:greenfunction}) is $g_\omega (\mathbf{r},\mathbf{s}) = - j k/(4 \pi) $ when $\mathbf{r} = \mathbf{s}$. 
% The pressure at the surface can be calculated only given the normal velocity at the same surface, by substituting (\ref{eq:greenfunction}) and (\ref{eq:euler}) into (\ref{eq:kh}), we get
% \begin{equation}
%     p(\mathbf{r}, \omega) = 
%     \begin{cases}
%          \displaystyle \int _{\mathcal{S}} -\frac{1}{2 \pi} \frac{\omega^2}{c}\rho_0 v_n(\mathbf{s}, \omega) \, d\mathcal{S},  & \text{if } \mathbf{r} = \mathbf{s}\\
%           \displaystyle \int _{\mathcal{S}} - \frac{1}{2 \pi} j\omega \rho _0 v_n (\mathbf{s}, \omega) \frac{e^{-j \frac{\omega}{c}||\mathbf{r}-\mathbf{s}||}}{||\mathbf{r}-\mathbf{s}||}  \, d\mathcal{S},  & \text{otherwise}
%         \end{cases}.
%     \label{eq:rayleigh}
% \end{equation}
% which is the so called Rayleigh's first integral.

The exterior normal velocity field can be calculated when given the normal velocity and pressure by simply applying Euler's equation (\ref{eq:euler}) at $\mathbf {r}$ to (\ref{eq:kh}) \cite{williams2000fourier}: 

\begin{equation} 
\begin{aligned} & v_n(\textbf{r},\omega) = \int_\mathcal{S} \bigg( \frac{1}{j\omega \rho_0} p(\textbf{s},\omega) \frac{\partial^2}{\partial \textbf{n}^2} g_\omega(\textbf{r},\textbf{s})  \\ & \qquad  - v_n(\textbf{s},\omega) \frac{\partial}{\partial \textbf{n}} g_\omega(\textbf{r},\textbf{s}) \bigg) d\mathcal{S}, \\ & \qquad \qquad \qquad \qquad \qquad \qquad 
 \text{where $\mathbf{r}$ is outside $\mathcal{S}$}. 
\end{aligned} \label{eq:khv} \end{equation}

Therefore, the sound field propagation can be characterized by KH integral and its variations (\ref{eq:kh2}), (\ref{eq:khv}), with the knowledge of the pressure and the normal velocity fields on an object's surface.

%===========================

%\section{}

\section*{Acknowledgment}
This work was partially supported by the European Union - Next Generation EU under the Italian National Recovery and Resilience Plan (NRRP), Mission 4, Component 2, Investment 1.3, CUP D43C22003080001, partnership on "Telecommunications of the Future" (PE00000001 - program ``RESTART'').

\ifCLASSOPTIONcaptionsoff
  \newpage
\fi



\bibliographystyle{IEEEtran}
\bibliography{references}
\end{document}